\documentclass{jfm}
\usepackage{graphicx}
\usepackage{epstopdf, epsfig}
\usepackage{stmaryrd}
\usepackage{pifont}
\usepackage{amsmath,amssymb} 

\shorttitle{Solutal Marangoni instability in layered flows}
\shortauthor{J. R. Picardo, Radhakrishna T.G. and S. Pushpavanam}

\title{Solutal Marangoni instability in layered two-phase flows}

\author{Jason R. Picardo, Radhakrishna T.G.
 \and S. Pushpavanam\corresp{\email{spush@iitm.ac.in}}}

\affiliation{Department of Chemical Engineering, Indian Institute of Technology Madras,
Chennai, TN 600036, India
}

\begin{document}

\maketitle

\begin{abstract}
In this paper, the instability of layered two-phase flows caused by the presence of a soluble surfactant (or a surface active solute) is studied. The fluids have different viscosities, but are density matched to focus on Marangoni effects. The fluids flow between two flat plates, which are maintained at different solute concentrations. This establishes a constant flux of solute from one fluid to the other in the base state. A linear stability analysis is performed, using a combination of asymptotic and numerical methods. In the creeping flow regime, Marangoni stresses destabilize the flow, provided a concentration gradient is maintained across the fluids. One long wave and two short wave Marangoni instability modes arise, in different regions of parameter space. A well-defined condition for the long wave instability is determined in terms of the viscosity and thickness ratios of the fluids, and the direction of mass transfer. Energy budget calculations show that the Marangoni stresses that drive long and short wave instabilities have distinct origins. The former is caused by interface deformation while the latter is associated with convection by the disturbance flow. Consequently, even when the interface is non-deforming (in the large interfacial tension limit), the flow can become unstable to short wave disturbances.  On increasing $\Rey$, the viscosity-induced interfacial instability comes into play. This mode is shown to either suppress or enhance the Marangoni instability, depending on the viscosity and thickness ratios. This analysis is relevant to applications such as solvent extraction in microchannels, in which a surface-active solute is transferred between fluids in parallel stratified flow. It is also applicable to the thermocapillary problem of layered flow between heated plates.
\end{abstract}


\section{Introduction}\label{sec:intro}

An instability of layered two-phase flows caused by inter-phase mass transfer of a surface active solute is investigated in this paper. This work is motivated by recent developments in microscale solvent extraction and two-phase heterogeneous reaction systems. Typically, two phases are brought into contact as parallel flowing streams in a microchannel. Mass transfer of a solute occurs between the phases, across a well defined interface. Such systems have been successfully applied to extract products from process streams and purify waste streams \citep{Assmann2013,Fries2008,hotokezaka}, separate biomolecules \citep{Znidarsic-Plazl2007} and carry out mass transfer limited heterogeneous reactions. An example of the latter process is phase transfer catalysis \citep{Sinkovec2013,Aljbour2010}, in which a catalyst facilitates the transport of reactant  and product species between immiscible phases. Layered or stratified two-phase flow in microchannels offers the advantages of small diffusion path length, high interfacial area to volume ratio, low shear rates and small inventories.

The physico-chemical behavior of these systems is complicated by the fact that the solute often behaves as a soluble surfactant, i.e. the interfacial tension of the two-fluid interface varies with the concentration of the solute. In such cases, changes in solute concentration along the interface generate Marangoni stresses that can significantly impact the flow, thereby modifying the rate of mass transfer. Examples of such solute/fluid-fluid systems, encountered in extraction processes, are acetone/water-toluene \citep{Javed1989}, Butyric acid/water-toluene \citep{sternling} and oxyethylated alcohols/water-heptane \citep{Tadmouri2010}. Several phase transfer catalysts also act as soluble surfactants and modify the interfacial tension of the interfluid interface \citep{Dutta1993}.

There are two fundamentally different routes through which Marangoni stresses may impact mass transfer in these flows. The first is due to the fact that the concentration at the interface generally varies significantly along the microchannel. The two phases enter the channel with a significant concentration difference and leave the channel close to or at equilibrium. Marangoni stresses due to this axial concentration gradient will certainly modify the steady state flow of the system. However, in a recent study, we have shown that this effect is not strong enough to impact the primary pressure driven flow, for practical fluid solute systems \citep{picardo}. The \textit{base steady state} flow remains almost the same as that in the absence of the solute. 

The second route through which Marangoni stresses can impact mass transfer is by generating a hydrodynamic instability that leads to a new, possibly dynamic, flow state. It is the aim of this work to establish if such an instability is possible, to determine the stability threshold and to understand the nature of the instability modes. Towards this end we study a model problem that is closely related to the extraction system, but in which the base steady flow is fully developed and free from Marangoni stresses. Specifically, we consider layered stratified flow between two infinite flat plates, which are maintained at different solute concentrations. The concentration gradient maintained across the fluids sustains mass transfer between the phases. The corresponding base steady flow is fully developed and unidirectional. We analyze the stability of this flow to infinitesimal perturbations via a classic normal mode analysis.

There is good reason to expect a Marangoni instability in this flow, on the basis of past work on the stability of stationary fluid layers sustaining mass transfer of a soluble surfactant. First analyzed by \citet{sternling}, this stationary \textit{solutal Marangoni} instability is the subject of a large body of literature. Much of this work is surveyed in the reviews by \citep{Schwarzenberger2014,Kovalchuk2006}. The stability is affected by the fluid and solute properties as well as the direction of mass transfer. The existence of a finite concentration gradient across the fluids is a necessary condition for instability of the stationary fluid layers.  Whether this remains a necessary condition when pressure driven flow is imposed is a question which we aim to answer in this paper.

A rather disjoint, though equally large body of literature exists on the stability of two-phase stratified Poiseuille flow \citep{boomkamp2}. In the absence of surfactants, this flow is unstable to a long wave interfacial instability called the viscosity-induced mode \citep{boomkamp2}. This mode was first identified by \citet{Yih1967} and occurs even at low Reynolds numbers. The flow is stable, however, in the creeping flow limit. At the other extreme of large Reynolds numbers, the shear mode becomes unstable \citep{boomkamp2,yiantsios}. This mode takes the form of Tollmein-Schilichting waves and is induced by Reynolds stresses near the walls. The shear instability is unimportant at the low Reynolds numbers encountered in milli and micro channels.

Studies on the effects of surfactants on two-phase stratified flow are largely limited to the case of insoluble surfactants. Frenkel and Halpern \citep{frenkel,halpern} first demonstrated that the otherwise stable creeping flow becomes unstable on introducing an insoluble surfactant. Further studies have explored the influence of inertia \citep{blythinertia,frenkelinertia} and nonlinear dynamics \citep{blythcreeping,Wei2005,Blyth2007,arghya}. From the perspective of this literature, the present study extends current understanding of surfactant effects to the case of soluble surfactants.

A few studies that consider the influence of shear flow on the solutal Marangoni instability have been carried out. \citet{sun} have analyzed the instability caused by mass transfer in gas-liquid Poiseuille flow. \citet{Zaisha2008} have carried out DNS simulations for the case of liquid-liquid Couette flow. Both studies are restricted to the case of a non-deforming interface. However, interfacial deformation has a profound impact on the solutal Marangoni instability when Poiseuille flow is present, as shown in this paper. Very recently \citet{You2014a} have presented stability results for the solutal Marangoni instability in Poisuelle flow. However, a key term that accounts for interfacial concentration gradients due to interface deformation is absent in their model. Here, we show that this term plays an important role in the Marangoni instability and cannot be ignored unless the interface is flat. 

We make certain assumptions that simplify the system while retaining its essential features. We consider two dimensional Poiseuille flow between flat walls of two immiscible, incompressible Newtonian fluids. The concentrations of the solute at the two walls are maintained at two different constant values. The channel width is assumed to be sufficiently small for buoyancy effects to be neglected in comparison with Marangoni effects. This occurs when the velocity scale of natural convection induced flow ($\beta_D \Delta C g d^2/\mu$) to Marangoni stress induced flow ($\beta \Delta C d/l_c \mu$) is small ($d g l_c \beta_D /\beta \ll 1$), for each fluid. Here, $\mu$  and $d$ are the fluid's viscosity and channel width respectively. $\beta_D$ and $\beta$ are measures of the sensitivity of density and interfacial tension to solute concentration respectively. $\Delta C$ is the concentration difference applied across the plates and $l_c$ is the longitudinal length scale of streamwise variations in concentration. As buoyancy effects are unimportant in this regime, we simplify the model by assuming the densities of the fluids to be equal and independent of concentration. 

We also make certain simplifying assumptions regarding the properties of the soluble surfactant. A linear dependence of interfacial tension on concentration is considered. The rate of solute adsorption/desorption to/from the interface is assumed to be instantaneous in comparison with transport processes on the interface and in the bulk. Under these conditions, the interface solute concentration will be nearly in equilibrium with the bulk concentration in the adjacent fluid on either side of the interface. The model can then be written entirely in terms of bulk phase concentrations. It consists of a solute transport equation in each fluid and two interface boundary conditions that enforce equality of flux and local equilibrium at the interface \citep{picardo}.

Apart from model simplification, an additional advantage of assuming the aforementioned solute properties is that the results and conclusions of this study can be directly applied to thermocapillary instabilities, which arise due to variations of temperature. In the thermal analogue of this problem, the walls are maintained at different temperatures and heat transfer occurs between the fluids. Marangoni stresses are generated due to the dependence of interfacial tension on temperature. The corresponding governing equations are identical to those considered in this paper, provided an appropriate interchange of physical quantities is made (e.g. replacing the solute diffusion coefficients by thermal diffusion coefficients of the fluids). \citet{sternling2} were among the first to study the thermocapillary instability in stationary fluid layers. \citet{gumerman1} have studied the stability of stratified \textit{Couette} flow between heated plates, simultaneously considering bouyancy and thermocapillary effects. However, the parameter space was not fully explored, as only a few specific fluid systems were analyzed. A detailed asymptotic analysis of thermocapillary instability in Couette flow was carried out by \citet{Wei2006} for the case of one fluid layer being much thinner than the other (thin-layer limit).

The outline of this paper is as follows. The governing equations are presented in \S \ref{sec:prob}. The steady base state concentration and velocity fields are presented in \S \ref{sec:base}. Brief descriptions of the linear stability analysis and energy budget calculation are presented in \S \ref{sec:LSA} and \S \ref{sec:energy}. An asymptotic analysis for long wave disturbances is carried out in \S \ref{sec:longwave}. The results of numerical calculations for all wave numbers are discussed next. \S \ref{sec:longandshort} to \S \ref{sec:flux} consider the limit of creeping flow, wherein  Marangoni effects are the only possible source of instability (the viscosity induced mode is stable for creeping flow).  In \S \ref{sec:longandshort} three different types of instability modes - one long wave and two short wave - are identified. Mode switching between the two short wave modes is studied in \S \ref{sec:comp}. The results therein reveal a key qualitative difference between the two short wave modes. The transition from long waves to short waves is analyzed in \S \ref{sec:trans}. These numerical results along with the long wave analysis of \S \ref{sec:longwave} points to a well defined transition boundary in parameter space. \S \ref{sec:flux} investigates whether a finite concentration difference across the fluids is a necessary condition for the Marangoni instability. The analysis is extended beyond the creeping flow limit to small but finite Reynolds numbers in \S \ref{sec:inertia}. This introduces the viscosity-induced mode in addition to the three previous Marangoni modes. The influence of inertia on the Marangoni modes, as well as the effect of a soluble surfactant on the viscosity induced mode are studied in this section. In \S \ref{sec:compare} we compare our results with previous work by \citet{Wei2006} and \citet{You2014a}. The key results and conclusions of this work are summarized in \S \ref{sec:conclusion}, along with some suggestions for future work.

\section{Governing Equations}\label{sec:prob}
\begin{figure}
  \centerline{\includegraphics[scale=0.74]{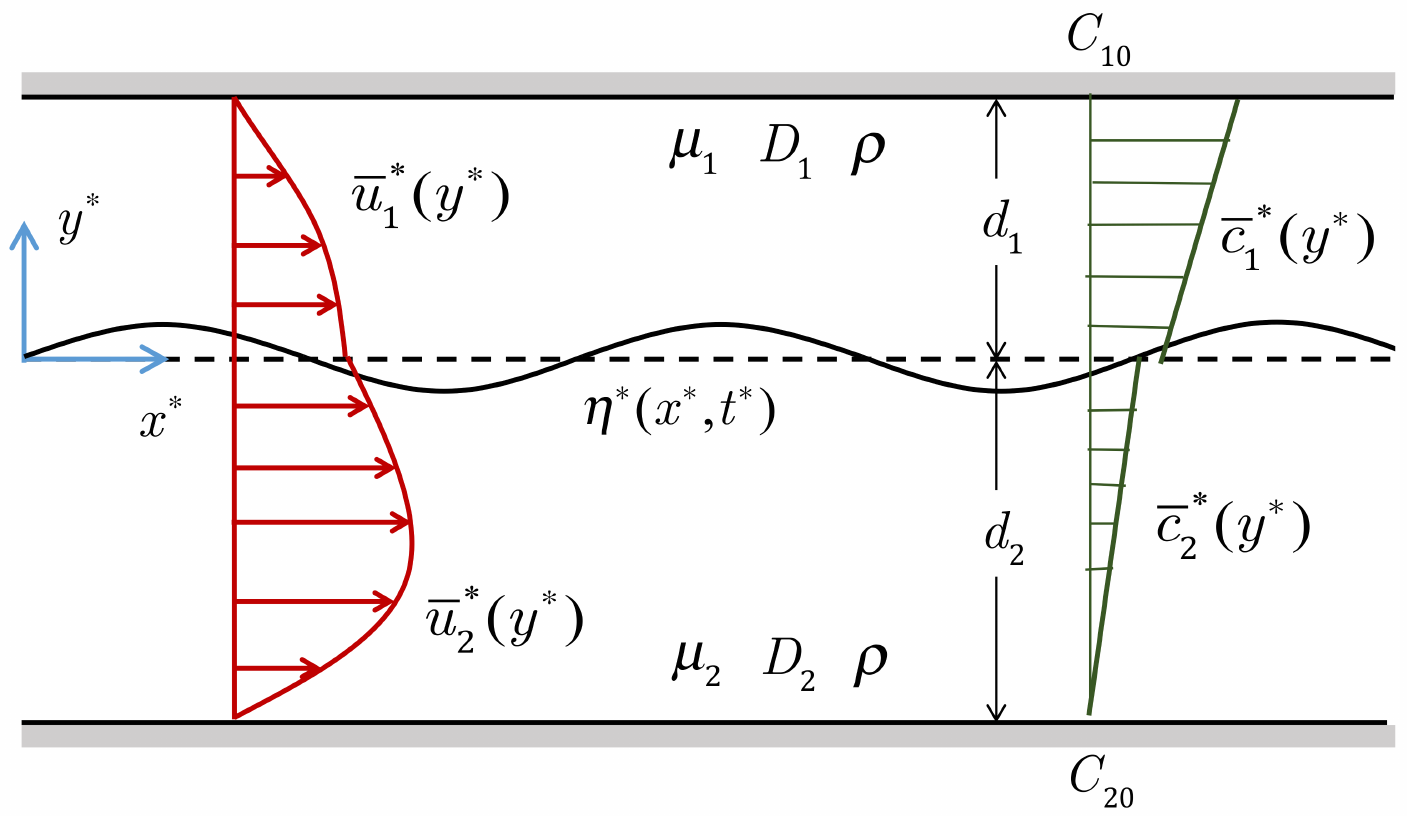}}
  \caption{Schematic of the system under study. The deformable interface is located at $y^*=\eta^*(x^*,t^*)$. The dashed line at $y^*=0$ corresponds to the undeformed interface in the base state. $\bar{u}^*_i$ and $\bar{c}^*_i$ are the velocity and concentration fields in the base state. The concentration field obeys the linear equilibrium condition $\bar{c}^*_1=K \bar{c}^*_2$ at the interface. Since $K$ can differ from unity, the concentration field can be discontinuous at the interface in general. Both fluids have a density $\rho$, while $\mu_i$ and $D_i$ are the viscosities and solute diffusivities respectively. }
\label{fig:schematic}
\end{figure}

A schematic of the 2D flow system being investigated is shown in Fig. \ref{fig:schematic}. Two immiscible, incompressible viscous fluids with equal densities flow side-by-side between two infinite parallel plates separated by $d_1+d_2$. The liquid-liquid interface is located at $y^*=0$ (superscript `$*$' indicates that the quantity is dimensional). The flow is driven by an imposed pressure gradient parallel to the $x^{*}$-axis. Concentrations at the top and bottom plates are maintained at $C_{10}$ and $C_{20}$, respectively.

Choosing the interfacial velocity ($U_0$), the thickness of the top fluid layer ($d_1$) and the concentration at the top plate ($C_{10}$) as the characteristic velocity, length and concentration scales, the dimensionless coordinate directions ($x,\,y$), streamwise velocity ($u_j'$), transverse velocity ($v_j'$), pressure ($p_j'$) and concentration ($c_j'$) fields are defined as 
\begin{equation}
	(x, y)=\frac{(x^*, y^*)}{d_1},\;(u_j', v_j')=\frac{(u_j^*, v_j^*)}{U_0},\;p_j'=\frac{p_j^*}{\mu_jU_0/d_1},\;c_j'=\frac{c_j^*}{C_{10}} \label{scaling}
\end{equation}
The subscript $j$ takes the values 1 and 2 to denote the top and bottom fluids, respectively. The dimensionless governing equations, viz. the continuity equation, Navier-Stokes equation and the species transport equation, are given by
\begin{subequations}
\begin{equation}
	\bnabla \bcdot \boldsymbol{v}_j'=0
\end{equation}
\begin{equation}
	\frac{\Rey}{m_j}\bigg(\frac{\partial \boldsymbol{v}_j'}{\partial t}+\boldsymbol{v}_j'\bcdot \bnabla \boldsymbol{v}_j'\bigg)=-\bnabla p_j'+\nabla^2\boldsymbol{v}_j'
\end{equation}
\begin{equation}
	\frac{\partial c_j'}{\partial t}+\boldsymbol{v}_j'\bcdot\bnabla c_j'=\frac{D_{r,j}}{\Pen}\nabla^2c_j'
\end{equation}
\label{governing}
\end{subequations}
with $j=1,\,m_1=1,\,D_{r,1}=1$ for fluid 1 and $j=2,\,m_2=m=\mu_2/\mu_1,\,D_{r,2}=D_r=D_2/D_1$ for fluid 2. Here,
\begin{equation}
	\bnabla=\bigg(\frac{\partial}{\partial x},\,\frac{\partial}{\partial y}\bigg);\;\boldsymbol{v}_j'=(u_j',\,v_j');\;\Rey=\frac{\rho U_0d_1}{\mu_1},\; \Pen=\frac{U_0d_1}{D_1}
\end{equation}
The dimensionless parameters $\Rey ,\,\Pen ,\,m$ and $D_r$ represent the Reynolds number, P\'eclet number, ratio of dynamic viscosities and molecular diffusivities, respectively.
All physical properties are assumed to be constant with the exception of interfacial tension, which depends on the concentration of the solute at the interface.
 
In general, a separate solute balance equation must be written at the interface, which accounts for changes in interface concentration due to surface convection and diffusion along the interface, as well as adsorption/desorption \citep{leal}. However, if the rates of adsorption and desorption are very high, then the interface concentration may be assumed to be in equilibrium with the concentration in the adjacent fluid on either side of the interface. Under these conditions, the interface solute balance reduces to two boundary conditions at the interface, $y=\eta' (x,t)$, which require the diffusive flux to be continuous and the bulk concentrations to be in equilibrium. For dilute solutions these boundary conditions read as \citep{picardo}:”
\begin{equation}
	-\eta'_x c_{1,x}'+c_{1,y}'=D_r(-\eta'_x c_{2,x}'+c_{2,y}')
	\label{fluxbalance}
\end{equation}
\begin{equation}
	c_1'=Kc_2'
	\label{inteq}
\end{equation} 
Here subscripts $x$ and $y$ indicate partial differentiation and $K$ is the distribution coefficient. Note that the unit normal to the interface is given by $(\boldsymbol{e}_y-\eta'_x\boldsymbol{e}_x) (1+{(\eta'_x)^2})^{-1/2}$, while the unit tangent is given by $(\eta'_x\boldsymbol{e}_y+\boldsymbol{e}_x) (1+{(\eta'_x)^2})^{-1/2}$. Here $\boldsymbol{e}_y$ and $\boldsymbol{e}_x$ are unit vectors in the $y$ and $x$ directions respectively.

The value of interfacial tension is calculated using the bulk concentration at the interface. Assuming a linear dependence of interfacial tension on solute concentration, we have: 
\begin{equation}
	\sigma^*=\sigma_0(1-\beta C_{10} (c_1'-c_r'))
	\label{empi}
\end{equation}
where $\beta=-(d\sigma^*/dc_1^*)/\sigma_0$ and $\sigma_0$ is the interfacial tension when the solute concentration at the interface equals the reference value $C_{10}c_r'$. A positive value of $\beta$ implies that interfacial tension decreases with increasing solute concentration. Due to the linear equilibrium relationship \eqref{inteq}, either $c_1'$ or $c_2'$ can be used to express the dependency of interfacial tension on concentration \citep{sternling}.

The variation of interfacial tension causes tangential Marangoni stresses at the interface, which are balanced by shear stresses. Using relation \ref{empi}, the tangential stress balance at the interface reads:
\begin{equation}
	\frac{1}{\big(1+(\eta'_x)^2\big)^{1/2}}\big\llbracket 2\eta'_xm_j\big(v'_{j,y}-u'_{j,x}\big)+\big(1-(\eta'_x)^2\big)m_j\big(u'_{j,y}+v'_{j,x}\big)\big\rrbracket_2^1-\frac{Ma}{\Pen}\big(c_{1,x}'+\eta'_x c_{1,y}'\big)=0
	\label{tanstress}
\end{equation}
where the jump operator is defined as $\llbracket g_j \rrbracket_2^1=g_1-g_2$. $Ma$ is the Marangoni number, given by $\sigma_0\beta C_{10} d_1/D_1\mu_1$. We have not included the concentration difference between the plates in the Marangoni number, to allow for the possibility that the interface sensitivity parameter ($\beta$) and the concentration difference (represented by $\gamma=C_{20}/C_{10}$) may have independent effects on the system's stability.

A similar description of solutal Marangoni effects has been used by \citet{sternling} and \citet{Smith1966} to study instabilities in stationary layered fluids. The two terms multiplied by $Ma$ in \eqref{tanstress} derive from the surface gradient of the solute concentration ($\bnabla_sc_1$). They account for variations of interfacial tension due to a non-uniform distribution of the solute along the interface. The first term represents concentration perturbations caused by the disturbance flow. The second term accounts for concentration variations that arise along a deformed interface, when the base concentration varies transversely. This latter term is absent in the analysis of \cite{You2014a}. 

Interfacial tension also exerts normal stresses which tend to maintain a flat interface. This effect is included in the normal stress balance at the interface, which reads:
\begin{gather}
	\big(1+(\eta'_x)^2\big)\big(mp_2'-p_1'\big)+2\big\llbracket(\eta'_x)^2m_j u'_{j,x}-\eta'_xm_j \big(u'_{j,y}+v'_{j,x}\big)+m_j v'_{j,y}\big\rrbracket_2^1
	\hspace{9em} \nonumber\\ \hspace{20em}
	= \; -\frac{1}{Ca}\bigg(\frac{\sigma^*}{\sigma_0}\bigg)\frac{\eta'_{xx}}{\big(1+(\eta'_x)^2\big)^{1/2}}
	\label{norstress}
\end{gather} 
$Ca=\mu U_0/\sigma_0$ is the dimensionless Capillary number, which represents the relative importance of capillary forces in comparison to viscous forces. Note that because the characteristic scale for pressure in each fluid contains the respective fluid's viscosity (cf. \eqref{scaling}), the dimensionless pressure is discontinuous across the interface even when the interface is flat with $v_i'=0$ ($mp_2'=p_1'$ in this case).

The remaining boundary conditions at the interface are the equality of velocities and the kinematic condition:
\begin{equation}
\boldsymbol{v}_1'=\boldsymbol{v}_2'
\label{continvel}
\end{equation}
\begin{equation}
\eta'_t+u_1'\eta'_x=v_1'
\label{kinematic}
\end{equation}
At the bounding plates, no-slip and no-penetration conditions are applied on the velocity field and Dirichlet conditions on the concentration field:
\begin{equation}
	u_1'=v_1'=0;\;c_1'=1 \quad\mbox{at\ }\quad y=1
\end{equation}
\begin{equation}
	u_2'=v_2'=0;\;c_2'=\frac{C_{20}}{C_{10}}=\gamma\quad \mbox{at\ }\quad y=-n
\end{equation}
where $n=d_2/d_1$ is the ratio of thickness of fluid layers.

Equations (\ref{governing}--\ref{kinematic}) govern the behavior of the layered flow system, accounting for inter-phase mass transfer of a soluble surfactant and associated Marangoni stresses. These equations extend the model used by \cite{sternling} to the case of layered Poiseuille flow. In this simplified model, the dissipative effect of surface viscosity is neglected in comparison with that of bulk viscosity. Studies on stationary fluid layers have found that surface viscosity has a stabilizing influence, but does not significantly modify the key features of the stationary Marangoni instability \citep{Hennenberg1977,Kovalchuk2006}. The model also neglects dynamic transport of the solute in the interface Gibbs adsorption layer. This idealization corresponds to the limit of instantaneous adsorption/desorption of the solute to and from the interface. As a result of this assumption, the present model \emph{cannot} be reduced to the case of an insoluble surfactant by taking the limit of zero solute diffusivity within the fluids. Insoluble surfactants are trapped at the interface and can only be transported along the interface \citep{frenkel}.

\section{Base state velocity and concentration profiles}\label{sec:base}
The steady base state consists of unidirectional fully developed flow with a flat inter-fluid interface. The corresponding concentration field is invariant along the flow direction. The transverse variation of the base state fields (denoted by an overbar), obtained by solving Eqs. \eqref{governing}-\eqref{kinematic} are given by:
\begin{subequations}
\begin{gather}
	\overline{u}_j=1+a_jy+b_jy^2,\quad \overline{v}_j=0\\
	\overline{c}_j=s_jy+t_j,\quad \mbox{for\ }j=1,2 \label{cbasestate}\\
	\mbox{with\ }a_1=\frac{m-n^2}{n(n+1)};\;b_1=-\frac{m+n}{n(n+1)};\;a_2=\frac{a_1}{m};\;b_2=\frac{b_1}{m};\nonumber\\
s_1=\frac{D_r(1-\gamma K)}{D_r+Kn};\;t_1=\frac{K(n+D_r \gamma)}{D_r+Kn};\;s_2=\frac{s_1}{D_r};\;t_2=\frac{t_1}{K}\nonumber
\end{gather}
\label{basestate}
\end{subequations}
The piecewise linear base concentration field has different slopes in each fluid, which depend on the diffusivity ratio and thickness ratio. The direction of mass transfer depends on $\gamma$, with $\gamma<1/K(>1/K)$ corresponding to mass transfer from plate 1 to plate 2 (plate 2 to plate 1). When the plates are maintained at equilibrium ($\gamma=1/K$), the concentration is constant in each fluid and no net mass transfer occurs.

Marangoni stresses are absent in the base state since the concentration is uniform along the interface. However, a perturbation to the base flow will disturb the uniformity of the concentration profile and lead to Marangoni stresses. These stresses may cause the initial perturbation to grow, resulting in an instability. 

In the next section we analyze the stability of this base state to infinitesimally small perturbations.

\section{Linearized equations}\label{sec:LSA}

Infinitesimally small perturbations (denoted by a hat) are imposed on the base state \eqref{basestate} as follows:
\begin{gather}
	\nonumber u_j'=\overline{u}_j(y)+\widehat{u}_j(x,y,t),\quad v_j'=\widehat{v}_j(x,y,t),\quad p_j'=\overline{p}_j(x)+\widehat{p}_j(x,y,t)\\
	c_j'=\overline{c}_j(y)+\widehat{c}_j(x,y,t),\quad \eta'=\widehat{\eta}(x,t),\quad j=1,2
	\label{perturbations}
\end{gather}
The disturbance velocity field can be expressed in terms of a disturbance streamfunction ($\widehat{\psi}_j$) as $\widehat{u}_j=\partial \widehat{\psi}_j/\partial y$ and $\widehat{v}_j=-\partial\widehat{\psi}_j/\partial x$.

We focus on a temporal stability analysis, in this work, and assume classic normal mode forms for the perturbations:
\begin{equation}
	\begin{bmatrix}
	\widehat{\psi}_j & \widehat{p}_j & \widehat{c}_j & \widehat{\eta}
	\end{bmatrix}=
	\begin{bmatrix}
	\psi_j(y) & p_j(y) & c_j(y) & h
	\end{bmatrix} \exp [{\mathrm{i}\alpha(x-\omega t)}]+\mbox{c.c.\ }
	\label{normalmodes}
\end{equation}
where c.c. denotes the respective complex conjugate and $\psi_j,\,p_j,\,c_j$ and $h$ are complex amplitudes of the corresponding normal modes. The streamwise wavenumber $\alpha$ is real while the wave speed $\omega$ is complex ($\omega=\omega_r+i\omega_i$). The growth rate of each normal mode is given by $\alpha \omega_i$. Hence, the system is unstable if $\omega_i>0$ for any $\alpha$.

The governing equations \eqref{governing} are linearized about the base state \eqref{basestate} to obtain evolution equations for the disturbance fields. Applying the stream function formulation and adopting the normal mode form of perturbations, we obtain:
\begin{subequations}
\begin{gather}
	\mathrm{i}\alpha \Rey \big[(\overline{u}_j-\omega)(\mathcal{D}^2-\alpha^2)\psi_j-2b_j\psi_j\big]=m_j(\mathcal{D}^2-\alpha^2)^2\psi_j \label{momgov}
	\\
	\mathrm{i}\alpha \Pen\big[(\overline{u}_j-\omega)c_j-s_j \psi_j\big]=D_{r,j}(\mathcal{D}^2-\alpha^2)c_j\label{congov}
\end{gather}
	\label{fullgoveq}
\end{subequations}
where the operator $\mathcal{D}$ refers to differentiation with respect to $y$ i.e. $\mathcal{D}=d/dy$. The boundary conditions at the interface are simplified for the case of small deflections ($h \ll 1$) using the domain perturbation technique \citep{Johns2002}. The resulting boundary conditions for the normal mode amplitudes are:
\begin{subequations}
\begin{gather}
	\psi_1(1)=\mathcal{D}\psi_1(1)=c_1(1)=\psi_2(-n)=\mathcal{D}\psi_2(-n)=c_2(-n)=0\\
	\mathcal{D}\psi_1(0)+ha_1=\mathcal{D}\psi_2(0)+ha_2\label{ujump}\\
	\psi_1(0)=\psi_2(0)\label{vcont}\\
	c_1(0)+hs_1=K(c_2(0)+hs_2)\label{cjump}\\
	\mathcal{D}c_1(0)=D_r\mathcal{D}c_2(0)\\
	(\mathcal{D}^2+\alpha^2)\psi_1(0)-m(\mathcal{D}^2+\alpha^2)\psi_2(0)=\mathrm{i}\alpha \frac{Ma}{\Pen}(c_1(0)+hs_1)\\
	m\mathcal{D}^3\psi_2(0)-\mathcal{D}^3\psi_1(0)-3\alpha^2(m\mathcal{D}\psi_2(0)-\mathcal{D}\psi_1(0))=(1/Ca)\,\mathrm{i}\alpha^3 h\\
	(\omega-1)h=\psi_1(0) \label{kinematiclinear}
\end{gather}
	\label{perbcs}
\end{subequations}
Here, the reference concentration at which interfacial tension takes the value $\sigma_0$ (cf. \eqref{empi}) is taken to be the interface concentration in the base state ($c_r'=\bar{c}_1(0)=t_1$). Note that \eqref{ujump} and \eqref{vcont} are obtained from the $x$ and $y$ components of the continuity of velocity condition \eqref{continvel}, while \eqref{kinematiclinear} is obtained from the kinematic equation \eqref{kinematic}.

Equations \eqref{fullgoveq} \& \eqref{perbcs} constitute a linear differential eigenvalue problem with $\omega$ as the eigenvalue. For the system to be stable, the entire spectrum of eigenvalues must lie in the lower half of the complex plane ($\omega_i<0$).

\section{Energy budget}\label{sec:energy}

To aid in the analysis of instability modes, we carry out an energy budget analysis. Following the procedure described in \citet{boomkamp2} and \citet{lin}, we take the inner product of the linearized momentum equations with the perturbation velocity vector ($\widehat{\boldsymbol{v}}_j$) and integrate across the transverse direction in each fluid. The Gauss divergence theorem is applied to the stress integrals and the entire equation is averaged over one axial wavelength $\lambda=2\upi/\alpha$ and one time period $T=2\upi/(\alpha \omega_R$). This results in the following mechanical energy balance equation:
\begin{gather}
	\sum_{j=1}^2KE_j=\sum_{j=1}^2REY_j+\sum_{j=1}^2DIS_j+NOR+TAN_{\mu}+MAS_I+MAS_F \label{mechenergy}
\end{gather}

where
\begin{gather}
	KE_j=\frac{\Rey}{T\lambda}\int_0^T dt\int_0^\lambda dx\int_{e_j}^{f_j}dy\bigg[\frac{d}{dt}\bigg(\frac{\widehat{u}_j^2+\widehat{v}_j^2}{2}\bigg)\bigg]\label{KE}\\
	REY_j=-\frac{\Rey}{T\lambda}\int_0^T dt\int_0^\lambda dx\int_{e_j}^{f_j}dy\bigg[\widehat{u}_j\widehat{v}_j\frac{d\overline{u}_j}{dy}\bigg]\\
	DIS_j=-\frac{m_j}{T\lambda}\int_0^T dt\int_0^\lambda dx\int_{e_j}^{f_j}dy\big[2 \widehat{u}_{j,x}^2+2 \widehat{v}_{j,y}^2+(\widehat{u}_{j,y}+\widehat{v}_{j,x})^2\big] \label{dis}\\
	NOR=\frac{1}{T\lambda Ca}\int_0^T dt\int_0^\lambda dx\big[\widehat{v}_1 \widehat{\eta}_{xx}\big]_{y=0}\\
	TAN_{\mu}=\frac{1}{T\lambda}\int_0^T dt\int_0^\lambda dx\big[(\widehat{u}_2-\widehat{u}_1)(\widehat{u}_{1,y}+\widehat{v}_{1,x})\big]_{y=0}\\
	MAS_I=-\frac{1}{T\lambda}\frac{Ma}{\Pen}\int_0^T dt\int_0^\lambda dx \bigg[ \widehat{u}_2 \widehat{\eta}_x \frac{d\overline{c}_1}{dy}\bigg]_{y=0}\\
	MAS_F=-\frac{1}{T \lambda}\frac{Ma}{\Pen}\int_0^T dt\int_0^\lambda dx\big[ \widehat{u}_2 \widehat{c}_{1,x}\big]_{y=0}
\end{gather}

Here $j=1,\,e_1=0,\,f_1=1$ for fluid 1 and $j=2,\,e_2=-n,\,f_2=0$ for fluid 2.

On the LHS of the mechanical energy balance is the sum of the disturbance kinetic energy of the two fluids. This term is positive when a finite $Re$ flow is unstable. On the RHS are six different work terms, which either produce kinetic energy or consume it. $\sum_{j=1}^2REY_j$ is the total energy transfered from the base state to the disturbance flow by Reynolds stresses. This term becomes important only at relatively high $\Rey$. The next term, $\sum_{j=1}^2DIS_j$ is always negative (cf. \eqref{dis}) and represents viscous dissipation. The other four terms correspond to work done by stresses at the interface. $NOR$ represents the contribution from normal Capillary forces while $TAN_\mu$ is associated with the viscosity difference between the fluids. When the viscosities are unequal, $\widehat{u}_2 \ne \widehat{u}_1$ (cf. \eqref{ujump}) and this term can transfer energy to the disturbance flow. It is responsible for the viscosity-induced instability, first identified by \citet{Yih1967}, that occurs at small but finite $\Rey$. The last two terms are associated with Marangoni stresses due to a non-uniform distribution of soluble surfactant at the interface. $MAS_I$ accounts for concentration variations caused by interface deformation when the base concentration field varies in the transverse direction. The transverse concentration gradient causes the crest and trough of a deformed interface to have different solute concentrations and hence different values of interfacial tension. The second Marangoni stress term - $MAS_F$ - is caused by concentration perturbations associated with the disturbance flow. This term will be present even if the interface is flat. These two mechanisms were identified by \citet{Goussis1990} in the context of thermocapillary instability of a flowing heated liquid film.

Analogous to the mechanical energy equation \eqref{mechenergy}, we derive a balance equation for a concentration energy functional in each fluid, defined as $\widehat{c}_j^2/2$. Because the kinetic energy is identically zero in creeping flow ($KE_j=0$), the presence of an instability cannot be detected from the mechanical energy equation. Instead the growth of the concentration energy functional can be used to identify an instability. In addition, this equation provides insight into the growth of concentration perturbations.  

This equation is derived by multiplying the linearized solute balance with $\widehat{c}_j$, integrating across the fluids and averaging over one axial wavelength $\lambda=2\upi/\alpha$ and one time period $T=2\upi/(\alpha \omega_R$). Using the Gauss divergence theorem, we obtain:
\begin{gather}
	\sum_{j=1}^2E_j^c=\sum_{j=1}^2DIF_j+\sum_{j=1}^2CONT_j+INT\label{concenergy}\\
	E_j^c=\frac{1}{T\lambda}\int_0^Tdt\int_0^\lambda dx\int_{e_j}^{f_j}dy\bigg[\frac{d}{dt}\bigg(\frac{\widehat{c}_j^2}{2}\bigg)\bigg]\\
	DIF_j=-\frac{D_{r,j}}{\Pen}\frac{1}{T\lambda}\int_0^Tdt\int_0^\lambda dx\int_{e_j}^{f_j}dy \big[\widehat{c}_{j,x}^2+\widehat{c}_{j,y}^2\big]\label{dif}\\
	CONT_j=-\frac{1}{T\lambda}\int_0^Tdt\int_0^\lambda dx\int_{e_j}^{f_j}dy\bigg[\widehat{c}_j\widehat{v}_j\frac{d\overline{c}_j}{dy}\bigg]\\
	INT=-\frac{1}{\Pen}\frac{1}{T\lambda}\int_0^Tdt\int_0^\lambda dx\big[(\widehat{c}_1-\widehat{c}_2)\widehat{c}_{1,y}\big]_{y=0}
\end{gather}

The dynamics of the concentration energy functional is governed by the three terms on the RHS. $DIF_j$ term is always negative (cf. \eqref{dif}) and represents the damping effect of diffusion. The other two terms can cause concentration disturbances to grow. $CONT_j$ is associated with bulk convection effects in the transverse direction. This term contributes only in the presence of a finite base state concentration gradient ($d\overline{c}_j/dy \ne 0$). $INT$ is associated with a jump in the disturbance concentration across the interface, which occurs if the solute has a greater affinity for one of the fluids ($K \ne 1$), or has a greater diffusivity in one of the fluids ($D_r \ne 1$), or both (cf. \eqref{cjump} and \eqref{cbasestate}).

These energy balance equations are used to get insight into the mechanism driving an instability. The eigenfunctions, calculated by solving equations \eqref{fullgoveq} and \eqref{perbcs}, are substituted into the energy equations. The terms with a large magnitude correspond to dominant effects. The sign of these terms show whether they are stabilizing (negative) or destabilizing (positive) influences. Therefore, identifying the largest positive work term in the energy equation helps to identify the source of the instability.

\section{Longwave asymptotic analysis}\label{sec:longwave}

In this section we obtain an asymptotic solution to the eigenvalue problem \eqref{fullgoveq}-\eqref{perbcs} in the limit of long waves ($\alpha \to 0$). In order to understand the interaction of inertial and Marangoni effects, we assume both $\Rey$ and $Ma$ to be $O(1)$. Other parameters $(m,n,k,D_r,Ca,\Pen,\gamma)$ are also assumed to be $O(1)$. The solution is obtained as a regular perturbation expansion in $\alpha$, which yields the following expression for $\alpha \omega$:

\begin{gather}
\alpha \omega= \alpha \bigg [ 1+\frac{2n(m-1)(m-n^2)}{m^2+4mn+6mn^2+4mn^3+n^4} \bigg]\;\;\;\;\;\;\;\;\;\;\;\;\;\;\;\;\;\;\;\;\;\;\;\;\;\;\;\;\;\;\;\;\;\;\;\;\;\;\;\;\;\;\;\;\;\;\;\;\;\;\;\;
 \nonumber\\
+\mathrm{i}\alpha ^2 \bigg[ \frac{Ma}{\Pen}\frac{D_r K n^2(n+1)(n^2 -m)(1-\gamma K)}{2(D_r+Kn)^2(m^2+4mn+6mn^2+4mn^3+n^4)}\;\;\;\;\;\;\;\;\;\;\; 
\nonumber\\ \;\;\;\;\;\;\;\;\;\;\;\;\;\;\;\;\;\;
-\Rey \frac{(n^2-m)(m-1) G(n,m)}{420 m^2 (1+n)^2 (m^2+4mn+6mn^2+4mn^3+n^4)^3} \bigg]+ O(\alpha^3) \label{longexp}
\end{gather}

The speed of a travelling wave mode of wavelength $\alpha$ is given by $\omega_r$ while its growth rate is given by $\alpha \omega_i$ (the imaginary part of \eqref{longexp}). The growth rate consists of two terms: the first is due to Marangoni effects while the second accounts for the viscosity induced mode that is active at non-zero $\Rey$. The latter term is the same as that obtained by \citet{yiantsios} (verified by reproducing Fig. 2a of their paper).  The expression for $G(n,m)$ is given in Appendix A. Numerical evaluation shows this function to be positive definite over a wide range of $n$ and $m$, spanning $(10^{-4},10^4)$ in both parameters.

Let us first focus on the case of creeping flow ($\Rey=0$) wherein only Marangoni effects are important. From \eqref{longexp} it is clear that long waves become unstable when $n^2>m$, provided mass transfer occurs from fluid 1 to fluid 2 ($\gamma < 1/K$) and interfacial tension decreases with increasing solute concentration ($Ma>0$, i.e. $\beta>0$). The magnitude of the growth rate is seen to increase with the sensitivity of interfacial tension to surfactant concentration (i.e. increases with $Ma$) and with the applied concentration difference across the plates ($1-\gamma K$).

Switching the direction of mass transfer (i.e. having $\gamma > 1/K$) inverts the condition for long wave instability to $n^2<m$. Thus for a given pair of fluids and thickness ratio, the direction of mass transfer decides whether a long wave instability is present or not. This relation between long wave instability and the direction of mass transfer will reverse in case the soluble surfactant has the relatively uncommon property of increasing interfacial tension by its presence at the interface ($Ma<0$) \citep{Harkins1916,Evans1937,Wang2013a}. In the remainder of this work, we assume that mass transfer occurs from phase 1 to phase 2 ($\gamma < 1/K$) and that $Ma>0$.

In case of the thermocapillary problem, interfacial tension always decreases with increasing temperature ($Ma>0$). Consequently, long waves will be unstable when $n^2>m$, provided plate 1 is hotter than plate 2.

Some insight into the role of the diffusivity ratio $D_r$ is provided by \eqref{longexp}. $D_r$ is seen to effect only the magnitude of the growth rate, but not its sign. Therefore, it plays no role in the condition for instability. This is in contrast to the case of solutal Marangoni instability in \textit{stationary} fluid layers, wherein the system can be unstable if mass transfer occurs from the fluid of lower diffusivity to that of the higher, even though the viscosity and thickness ratios are unity ($n^2=m=1$) \citep{sternling,Schwarzenberger2014}.

The maximum growth rate occurs at $D_r=Kn$. These properties correspond to a base state concentration profile in which the interface concentration is a weighted mean of the wall concentrations: $\bar{c_1}(0)=K \bar{c_2}(0))=(C_{10}+KC_{20})/2$. In the thermocapillary problem $K=1$ and the base state interface temperature for maximum instability is the arithmetic mean of the wall temperatures.

When $\Rey$ is non-zero, the growth rate is affected by the presence of the viscosity-induced mode that is associated with a discontinuity of the slope of the base state velocity profile at the interface \citep{yiantsios,boomkamp2}. This discontinuity occurs when the viscosities are unequal ($m \ne 1$), provided the shear rate is non-zero at the interface ($n^2 \ne m$). Equation \eqref{longexp} shows that this mode can either counteract the longwave Marangoni instability or support it, depending on the viscosity ratio. For example, if $\gamma <1/K$, $n^2>m$ and $Ma>0$, then the Marangoni instability is suppressed (enhanced) on increasing $\Rey$ if $m>1$ ($m<1$). In such a case, the onset of long wave instability will occur only if $Ma$ is increased beyond a non-zero critical value. This critical threshold will increase with $\Rey$. On the other hand, if $n^2<m$ ($m>1$) then the viscosity induced mode will cause the growth rate to increase while Marangoni effects exert a stabilizing influence. A similar interaction between long wave Marangoni and viscosity-induced modes was identified in thermocapillary Couette flow by \citet{Wei2006}.

\section{Numerical Solution}\label{sec:numerics}
The eigenvalue problem (\eqref{fullgoveq}-\eqref{perbcs}) is solved numerically for arbitrary wavenumbers using the Chebyshev spectral collocation method described by \citet{boomkamp}. The dependent variables, $\psi_j$ and $c_j$, are expressed as series expansions of Chebyshev polynomials. Equations for the coefficients of the expansions are obtained by collocation of the governing equations on the interior points of a Gauss-Lobatto grid. Boundary conditions are applied at the boundary nodes. The result is a generalized $(N \times N)$ matrix eigenvalue problem of the form $\mathsfbi{A}\boldsymbol{x}=\omega \mathsfbi{B}\boldsymbol{x}$. Here $\boldsymbol{x}$ is vector of $N$ unknowns, which includes the $(N-1)/4$ coefficients in each Chebyshev polynomial expansion (for $\psi_1, \psi_2, c_1$ and $c_2$) and the amplitude of interface deformation $(h)$. The matrix $\mathsfbi{B}$ is singular due to the presence of zero rows (M in number) arising from the boundary conditions that do not contain $\omega$. Following \citet{boomkamp}, we use the corresponding $M$ equations to solve for $M$ unknown coefficients and replace them in the remaining $N-M$ equations. The result is a $(N-M) \times (N-M)$ generalized eigenvalue problem $\mathsfbi{A}'\boldsymbol{x}'=\omega \mathsfbi{B}'\boldsymbol{x}'$ in which $\mathsfbi{B}'$ is invertible. The eigenvalue spectrum is obtained using the QZ algorithm. To obtain $\psi_j$ and $c_j$, the eigenvectors $\boldsymbol{x}'$ must be transformed back to $\boldsymbol{x}$.

The computation is simplified in case of creeping flow ($\Rey=0$) because the momentum equations \eqref{momgov} simplify to
\begin{equation}
	(\mathcal{D}^2-\alpha^2)^2\psi_j=0,\quad \mbox{for\ }j=1,2
	\label{creeping}
\end{equation}
These equations \eqref{creeping} can be solved analytically to yield:
\begin{gather}
	\psi_1=\theta_1 \cosh (\alpha y)+\theta_2 \sinh (\alpha y)+\theta_3 y\cosh (\alpha y)+\theta_4 y\sinh (\alpha y)\\
	\psi_2=\theta_6 \cosh (\alpha y)+\theta_7 \sinh (\alpha y)+\theta_8 y\cosh (\alpha y)+\theta_9 y\sinh (\alpha y)
\end{gather}
where $\theta_j$ are arbitrary constants of integration. These expressions are substituted into the concentration equations \eqref{congov}, which are solved by the aforementioned Chebyshev spectral method. 

\begin{figure}
\centerline{\includegraphics[scale=1]{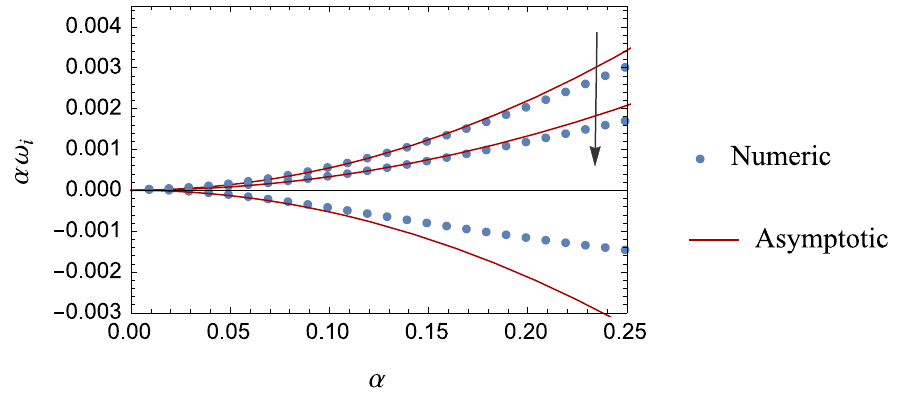}}
  \caption{Comparison of the dispersion curve obtained numerically using the spectral method with the long wavelength asymptotic relation \eqref{longexp}. Three cases, corresponding to $\Rey$ = 0, 10, 50 are depicted. The arrow indicates the direction of increasing $\Rey$. Other parameter values: $Ma$ = 10, $Ca$ = 1, $m$ = 1.5, $n$ = 2, $D_r$ = 0.5, $K$ = 0.5, $\Pen$ = 2, $\gamma$ = 0.5.}
\label{fig:asympnum}
\end{figure}

In the limit of $Ma \to 0$, the current system reduces to plane Poiseuille flow, the stability of which has been studied by \citet{yiantsios}. We have verified our numerical solution in this limit by comparing with their results. We have also compared our numerical predictions with the long wave asymptotic expansion \eqref{longexp}. The two are in good agreement in the limit of small $\alpha$, as shown in Fig. \ref{fig:asympnum}. In this figure the long wave Marangoni instability is stabilized as $\Rey$ is increased. This is in accordance with the prediction of \eqref{longexp} for the case of $n^2>m$ and $m>1$.

\section{Instabilities in creeping flow: Long wave and short wave modes}\label{sec:longandshort}

From this section onwards, we analyze the stability of the system to perturbations of all wavelengths, using the spectral collocation method. The range of values for $Ma$ and $\Pen$ selected for numerical calculations are appropriate to layered microchannel flows. In microchannels with dimensions of 50-200 $\mu$m, the flow rates of layered flow ranges from 20-150 $\mu$L/min \citep{hotokezaka,Znidarsic-Plazl2007,Fries2008}. Considering typical values of solute diffusivity of $O(10^{-9})$ m$^2$/s and liquid viscosity of $O(10^{-3})$ Pa$\,$s, we obtain $\Pen$ of $O(10^3)$ to $O(10^4)$. The magnitude of the variation of interfacial tension with concentration ($\sigma_0 \beta$) is typically of $O(10^{-6})$ Nm$^2$/mol \citep{sternling}. This results in $Ma$ of $O(10^3)-O(10^4)$. The value of $Ca$ is varied over a wide range to uncover all instability modes of the system and understand their behaviour. $Ca$ in microchannels however is rather small, of $O(10^{-3})-O(10^{-1})$. 

$\Rey$ for layered microchannel flows ranges from $O(1)-O(10)$, which is small, but not negligible. Nevertheless, we first analyse the case of creeping flow ($\Rey=0$), wherein Marangoni effects are the only source of instability. After understanding the pure Marangoni instabilities, the influence of small but finite $\Rey$ is examined. We begin in this section by identifying different types of Marangoni instability modes, in the limit of creeping flow, and study their characteristic features.

The numerical solution of the linear stability equations \eqref{fullgoveq} \& \eqref{perbcs} yields a spectrum of eigenvalues for each wavenumber $\alpha$. Of these, only two eigenvalues that have the largest and second largest imaginary parts play a role in deciding the system's stability. As $\alpha$ is varied these two eigenvalues trace out separate dispersion curves, which we label M1 and M2. Depending on parameter values, these two branches of eigenvalues can attain positive growth rates and render the system unstable. The real parts of the eigenvalues are always non-zero indicating that all modes have the form of travelling waves.

\begin{figure}
  \centerline{\includegraphics[scale=1]{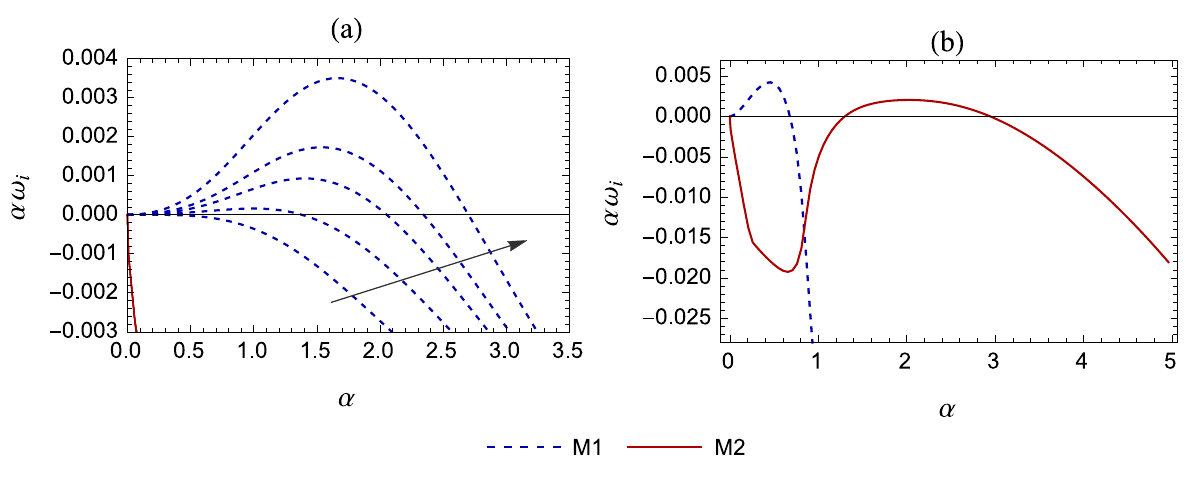}}
  \caption{Dispersion curves for the case of $n^2>m$, wherein long waves are unstable (a) Mode M1 manifests as a long wave instability. Dispersion curves are shown for a series of Marangoni numbers: 1, 1372.652, 3372.652, 5372.652, 11372.652. The arrow indicates the direction of increasing $Ma$. Other parameter values: $Ca$ = 100, $m$ = 1.5, $n$ = 1.24, $D_r$ = 0.5, $K$ = 0.5, $\Pen$ = 2000, $\gamma$ = 0.5. (b) Simultaneous instability of M1 long waves and M2 short waves. Parameter values: $Ma$ = 10000, $Ca$ = 1, $m$ = 0.9, $n$ = 1.3, $D_r$ = 0.5, $K$ = 1.2, $\Pen$ = 1000, $\gamma$ = 0.}
\label{fig:long}
\end{figure}

The asymptotic analysis for $\alpha \to 0$ in \S \ref{sec:longwave} (cf. \eqref{longexp}) revealed that the system is unstable for small $\alpha$ when $n^2 > m$ (provided mass transfer occurs from plate 1 to 2 or $\gamma < 1/K$). Dispersion curves for such a case are plotted in Fig. \ref{fig:long}a. Five different values of $Ma$ are considered in this figure. The M1 branch is seen to be unstable for a range of wavenumbers that extend from zero to some positive wavenumber $\alpha_0$. As $Ma$ is decreased to zero, the range of unstable wavenumbers shrinks and $\alpha_0 \to 0$. Based on these features, this mode is termed a \emph{long wave} (LW) instability. In this figure, the M2 branch is seen to be stable for all $\alpha$. This indicates that the long wave analysis of \S \ref{sec:longwave} only describes the asymptotic behavior of the M1 branch of eigenvalues and not the M2 branch. 

In all our numerical calculations, the M2 branch is found to be stable for $\alpha$ near zero. On the other hand, it can become unstable for a range of relatively large wavenumbers that are bounded away from zero. Such a case is shown in Fig. \ref{fig:long}b. In this figure, the M1 mode is unstable to long waves, while the M2 mode is unstable to \emph{short waves} (SW).

\begin{figure}
  \centerline{\includegraphics[scale=1]{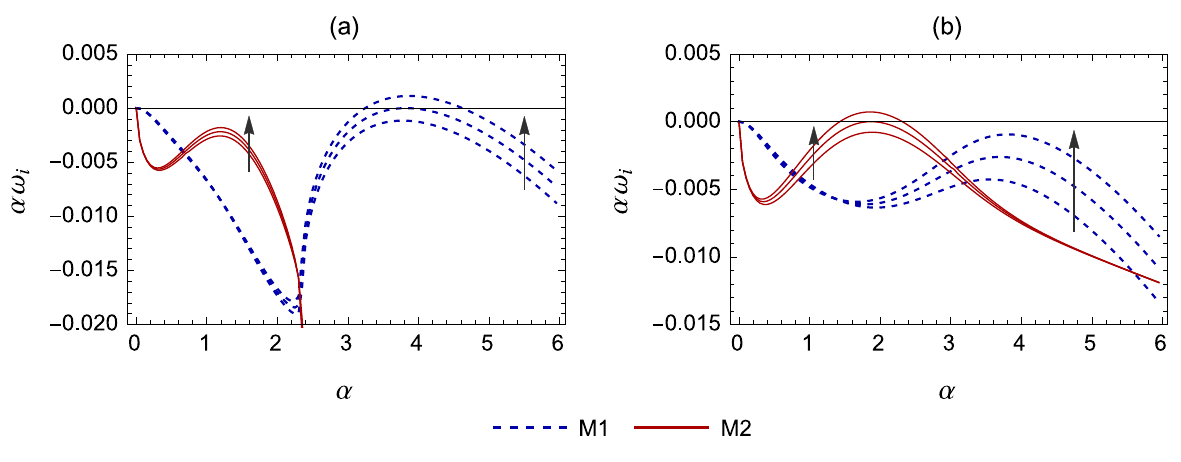}}
  \caption{Two instability modes M1 and M2 manifest as short wave instabilities beyond a critical Marangoni number. (a) Mode M1 becomes unstable; $Ca$ = 10, $Ma$ = 11631.21, 12231.21, 12831.21. (b) Mode M2 becomes unstable; $Ca$ = 100, $Ma$ = 9932.65, 10832.65, 11732.65. Other parameter values: $m$ = 1.5, $n$ = 1, $D_r$ = 0.5, $K$ = 0.5, $\Pen$ = 2000, $\gamma$ = 0.5. The arrow indicates the direction of increasing $Ma$.}
\label{fig:short}
\end{figure}

Next, we consider the case of $n^2<m$, for which the system is stable to long waves ($\alpha \to 0$), according to the asymptotic analysis (cf. \eqref{longexp}). In this case, only SW modes can become unstable.  They can originate from either the M1 branch or the M2 branch, provided $Ma$ is sufficiently large.  Fig. \ref{fig:short}a depicts a case in which the system just becomes unstable to M1 short waves, as $Ma$ is increased. Fig. \ref{fig:short}b shows the alternate case, wherein M2 short waves become unstable. As is apparent in both figures, SW modes become unstable only when $Ma$ is increased beyond a positive critical value. Beyond this value, a range of wavenumbers, bounded away from zero, become unstable. Typically, the range of unstable wavenumbers grows as $Ma$ is increased, as seen in Fig. \ref{fig:short}.

\begin{figure}
  \centerline{\includegraphics[scale=1]{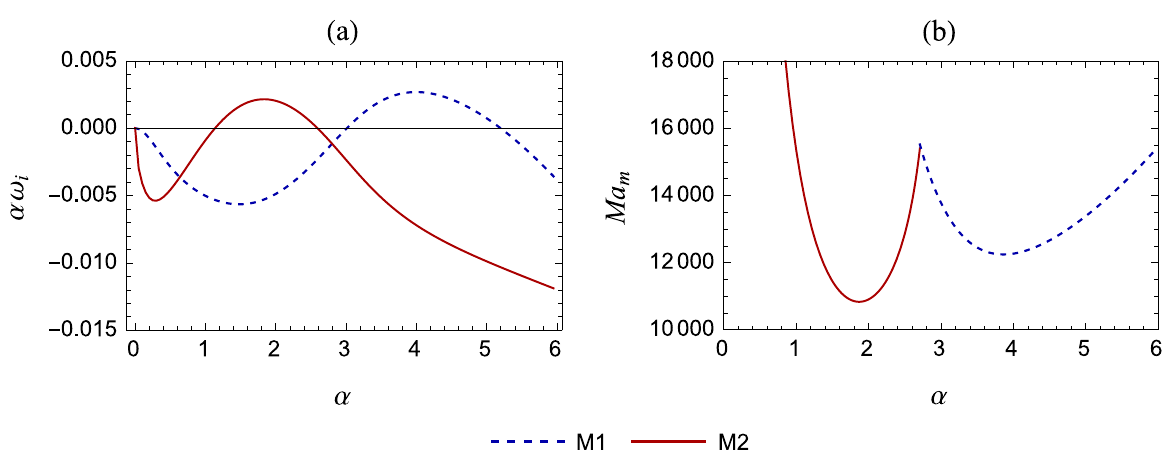}}
  \caption{(a) Dispersion curve showing instability to both short wave modes simultaneously ($Ma$ = 13732.65). (b) Neutral stability curve. Parameter values are given in the caption of Fig. \ref{fig:short}b.}
\label{fig:shortboth}
\end{figure}

In Figs. \ref{fig:short}a and \ref{fig:short}b, only one short wave mode is unstable in each case. However, the other short wave is not far from being unstable. In fact, if $Ma$ is increased to larger values, then both short wave modes can become unstable simultaneously. An example of this is shown in Fig. \ref{fig:shortboth}a. The parameters here are the same as Fig. \ref{fig:short}b, except that $Ma$ is larger. 

The dispersion curves plotted in Figs. \ref{fig:short}a, \ref{fig:short}b and \ref{fig:shortboth}a are useful for distinguishing the type of instability mode and for comparing their relative growth rates when $Ma$ is super-critical. However, the condition for stability of the system is more concisely represented by a neutral stability diagram. Fig \ref{fig:shortboth}b depicts such a diagram for the case studied in Figs. \ref{fig:short}b and \ref{fig:shortboth}a. This figure is obtained by determining the value of $Ma$ at which the growth rate of the largest eigenvalue is zero (called the marginal Marangoni number $Ma_m$), as a function of the wavenumber. The system is stable below this curve and unstable above it. The curve has two local minima, corresponding to the two short wave modes. The global minima of $Ma_m$ is the critical Marangoni number ($Ma_c$), above which the system first becomes unstable. In Fig. \ref{fig:shortboth}b, $Ma_c=10832.65$ and corresponds to the M2-SW mode. This implies that the M2-SW is critical at the onset of instability (as seen in Fig. \ref{fig:short}b). $Ma$ must be increased to supercritical values beyond 12500 for the M1-SW to become unstable as well.

The two short wave instability modes (Fig. \ref{fig:short}) differ from the long wave instability (Fig. \ref{fig:long}) not only in the qualitative nature of their dispersion curves, but also in their mechanical energy budgets. Table \ref{tab:short_long} presents the energy analysis for a long wave mode from Fig. \ref{fig:long} (corresponding to the fastest growing mode at $Ma = 3732.652$), the critical M1-SW mode from Fig \ref{fig:short}a and the critical M2-SW mode from Fig \ref{fig:short}b. In all cases, dissipation due to viscous forces is balanced primarily by Marangoni stress terms, affirming the fundamental association of these instabilities with solutal Marangoni forces. However, in case of long waves, $MAS_I$ is the dominant positive work term, whereas $MAS_F$ is the dominant positive term for both SW modes. Thus, the long wave instability is caused by concentration variations due to a deforming interface while the short wave instabilities are caused by concentration perturbations associated with the disturbance flow.

In summary, the system is susceptible to three different types of instability modes in the creeping flow limit: M1 long waves and M1 and M2 short waves. The M1 branch exhibits long waves if $n^2 > m$, and short waves otherwise. The M2 branch on the other hand only exhibits a short wave instability. However, these short wave modes can become unstable at any value of $m$ and $n$, if $Ma$ is increased sufficiently (cf. Fig. \ref{fig:long}(b) for $n^2>m$ and Fig \ref{fig:short}(b) for $n^2<m$). 

Long and short wave creeping flow instabilities also occur in the presence of an insoluble surfactant that is restricted to the inter-fluid interface \citep{halpern}. However, only two instability modes are present in that case - one long wave and one short wave. Both modes belong to the same branch of eigenvalues. The dynamics are therefore richer in the present case of a soluble surfactant. The additional short wave mode from the M2 branch introduces the possibility of instability to long and short wave modes simultaneously (eg. Fig. \ref{fig:long}b). Such a scenario is not observed when the surfactant is insoluble.

\begin{table}
  \begin{center}
\def~{\hphantom{0}}
  \begin{tabular}{lcccccccc}
      Mode  & $\alpha$  & $Ma$   &   $\sum_{j=1}^2DIS{_j}$ & $NOR$ & $TAN_{\mu} $ & \textbf{\textit{$MAS_{I}$}} & \textbf{\textit{$MAS_{F}$}} \\[3pt]
       M1-LW   & 1.410  & 3732.652 & -1 & $-3.30 \times 10^{-4}$ & $-3.94 \times 10^{-3}$ & 1.137 & -0.132\\
       M1-SW   & 3.800  & 12231.212 & -1 & $-1.50 \times 10^{-12}$ & $-7.9 \times 10^{-4}$ & -0.031 & 1.032\\
       M2-SW   & 1.851  & 10832.652 & -1 & $-1.66 \times 10^{-8}$ &0.105 & -0.077 & 0.973\\
  \end{tabular}
  \caption{Mechanical energy budget for LW and SW modes in Figs. \ref{fig:long} and \ref{fig:short}. The values have been normalized by the magnitude of total dissipation $\sum_{j=1}^2DIS{_j}$. The LW and SW have different energy signatures with the dominant positive work term being the $MAS_I$ in case of long waves and $MAS_F$ in case of short waves. $KE_i$ and $REY_i$ are identically zero in the creeping flow limit. Parameter values are as given in the captions of Figs. \ref{fig:long} and \ref{fig:short}.}
  \label{tab:short_long}
  \end{center}
\end{table}

\section{Switching between short wave modes}\label{sec:comp}

\begin{figure}
  \centerline{\includegraphics[scale=1]{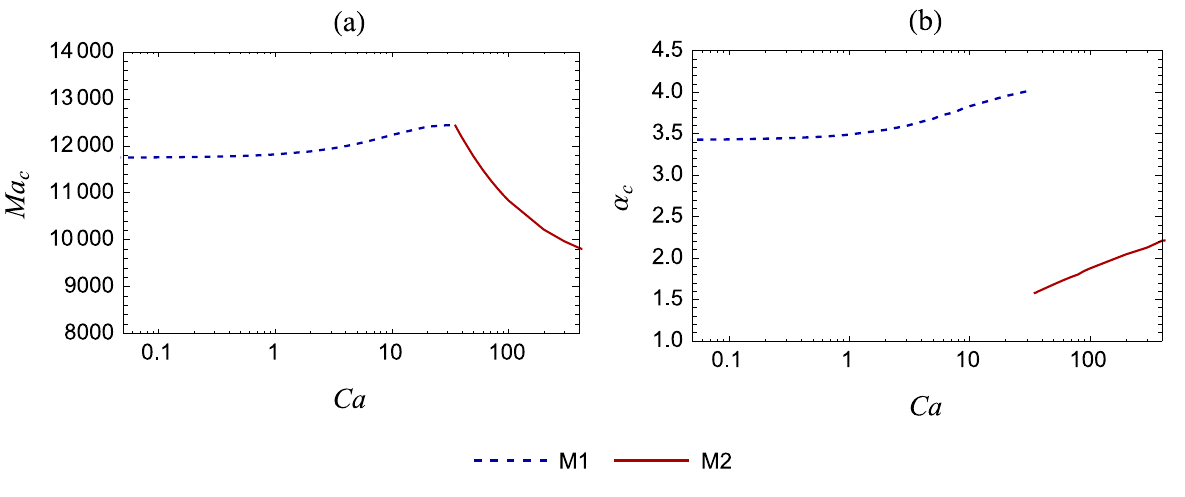}}
  \caption{Switching between short wave modes M1 and M2 as $Ca$ is varied. (a) Plot of the critical Marangoni number ($Ma_c$) (b) Plot of the critical wave number (${\alpha}_c$). Parameter values: $n$ = 1, $m$ = 1.5, $D_r$ = 0.5, $K$ = 0.5, $\Pen$ = 2000, $\gamma$ = 0.5.}
\label{fig:comp}
\end{figure}

\begin{figure}
  \centerline{\includegraphics[scale=1]{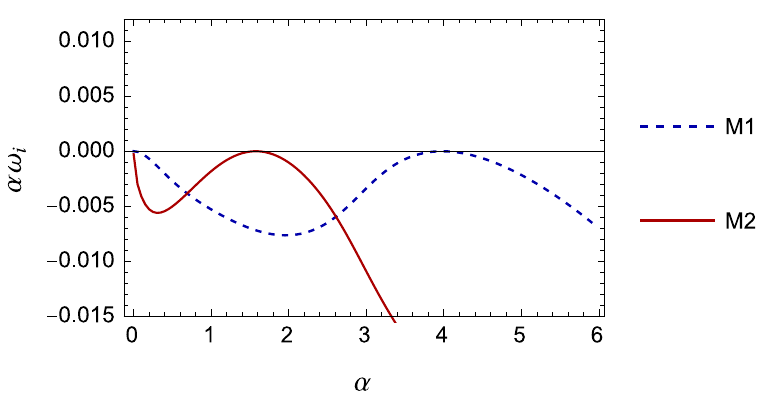}}
  \caption{Dispersion curves at the bi-critical point corresponding to the transition from critical M1-SW modes to critical M2-SW modes (codimension two bifurcation point). 
Parameter values: $m$ = 1.5, $n$ = 1, $Ca$ = 35, $Ma$ = 12430.34, $K$ = 0.5, $\Pen$ = 2000, $\gamma$ = 0.5.}
\label{fig:disp_comp}
\end{figure}

In this section we focus on the M1 and M2 short wave modes which can be simultaneously unstable when $n^2 < m$. Fig. \ref{fig:short} in \S \ref{sec:longandshort} demonstrated that the M1-SW mode is critical for small $Ca$ while the M2-SW mode is critical for large $Ca$ ($\mu U_0/\sigma_0$). To study the transition of the critical mode with $Ca$, we plot the critical Marangoni number ($Ma_c$), at which the system first becomes unstable, and the wavenumber of the corresponding critical mode ($\alpha_c$) as a function of $Ca$ in Fig. \ref{fig:comp}. Here, the critical mode is seen to switch from the M1-SW mode to the M2-SW mode as $Ca$ is increased. At the transition point, located at $Ca = 35$, both modes are critical, as shown in Fig. \ref{fig:disp_comp}. Across this point there is a jump in the critical wavenumber (cf. Fig. \ref{fig:comp}b). This point is a codimension-two bifurcation point at which the nature of the mode at the onset of instability changes abruptly.

The transition between modes is caused by the significantly different effects of $Ca$ on these modes, as demonstrated by Fig. \ref{fig:comp}. Decreasing $Ca$ strongly stabilizes the M2-SW mode while it has only a weak influence on the M1-SW mode. In fact, as $Ca \to 0$, the M1-SW mode becomes invariant to $Ca$. Since large values of interfacial tension (small $Ca$) prevent interface deformation, Fig. \ref{fig:comp} suggests that interfacial deformation plays an important role in the M2-SW mode but not in the M1-SW mode.

\begin{figure}
  \centerline{\includegraphics[scale=1]{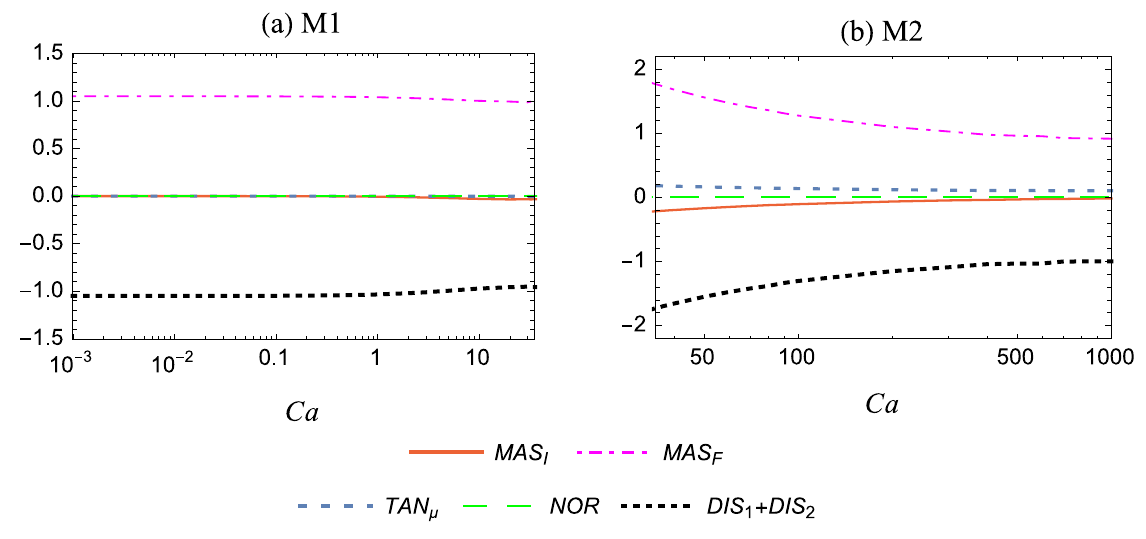}}
  \caption{Energy budget of the critical M1 and M2 short wave modes plotted in Fig. \ref{fig:comp} and Fig. \ref{fig:comp_Pe}b as a function of $Ca$. The $REY_i$ terms are identically zero because $\Rey=0$. Parameter values: $n$ = 1, $m$ = 1.5, $D_r$ = 0.5, $K$ = 0.5, $\Pen$ = 2000, $\gamma$ = 0.5.}
\label{fig:energyshort_Pe2000}
\end{figure}

To verify this hypothesis, an energy budget analysis of the critical modes is carried out for each value of $Ca$ in Fig. \ref{fig:comp}. The results are depicted in Figs. \ref{fig:energyshort_Pe2000}a and  \ref{fig:energyshort_Pe2000}b, for the M1-SW and M2-SW critical modes respectively. The dissipation is primarily balanced by the $MAS_F$ term, which is characteristic of both SW modes (cf. \S \ref{sec:longandshort}). The contribution from $NOR$ is also insignificant in both cases. The key difference between the two modes lies in the contributions of $MAS_I$ and $TAN_{\mu}$, which are non-zero only when the interface deforms. In case of the M1-SW mode, both terms are insignificantly small. On the other hand, their values are finite for the M2-SW mode, and grow larger as the transition point $Ca = 35$ is approached. These results imply that interfacial deformation is significant in the M2-SW instability mode, but not in the M1-SW mode. Consequently, on increasing interfacial tension, the M2-SW mode is stabilized as interface deformation is suppressed, whereas the M1-SW mode remains unstable.

\begin{figure}
  \centerline{\includegraphics[scale=1]{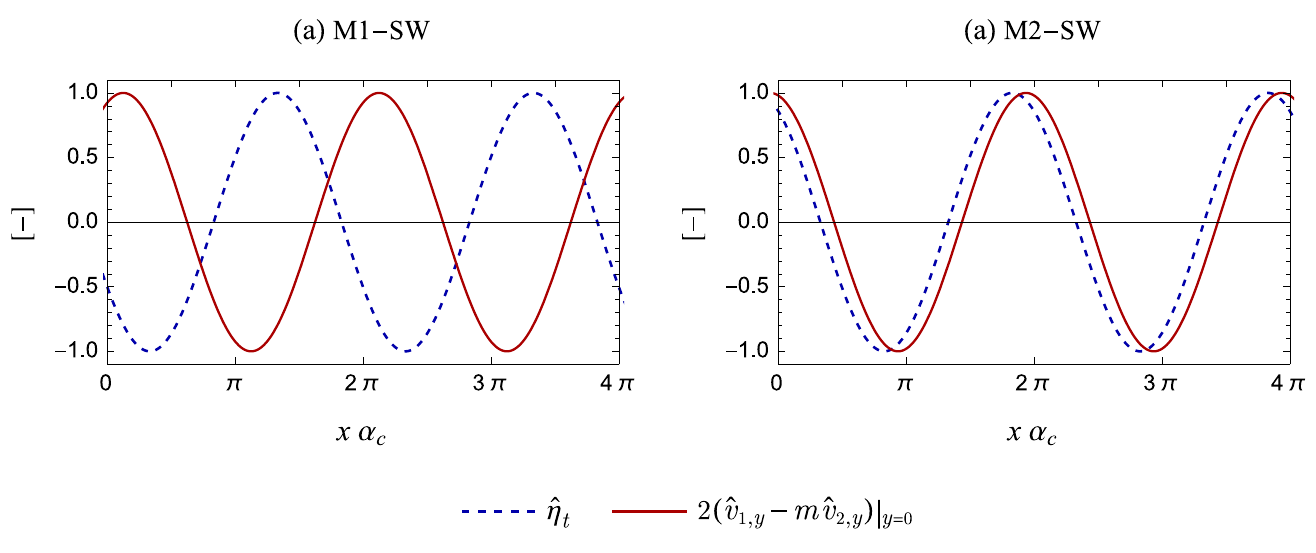}}
  \caption{Variation of normalized viscous normal stress along the interface, plotted along with the normalized local velocity of the interface, at leading order. (a) critical M1-SW mode (b) critical M2-SW mode. Parameter values are same as Fig. \ref{fig:disp_comp}, which corresponds to the co-dimension two bifurcation between M1-SW and M2-SW modes.}
\label{fig:intdefM1M2}
\end{figure}

This difference between the two modes is explained by the impact of the disturbance flow on the motion of the interface. Each instability mode introduces vertical fluid motion which exerts a net viscous normal stress on the interface. This is given by $2 \big( v_{1,y}-m v_{2,y}\big)|_{y=0}$, to leading order. To check whether this stress supports or counter-acts interface deformation, the phase difference ($\Delta \phi$) between the viscous normal stress and the velocity of the interface ($\widehat{\eta}_t$) is computed for both modes. For the case wherein both modes are critical (depicted in Fig. \ref{fig:disp_comp}), we find that the viscous normal stress is almost out-of-phase with interface velocity, in case of the the critical M1-SW mode ($\Delta \phi=0.79 \pi$), but nearly in-phase in case of the critical M2-SW mode ($\Delta \phi=0.10\pi$). The corresponding plots of normalized viscous stress and interface velocity are presented in Fig. \ref{fig:intdefM1M2}. This result implies that the disturbance flow of the M1-SW mode \textit{counteracts} interface deformation by exerting a downward stress on the interface, at locations where it is rising. In contrast, the disturbance flow of the M2-SW mode \textit{supports} interface deformation by exerting an \textit{upward} stress at positions where the interface is rising. Therefore, the M1-SW mode manifests without significant interface deformation, while the M2-SW mode is associated with a deforming interface.

\begin{figure}
  \centerline{\includegraphics[scale=1]{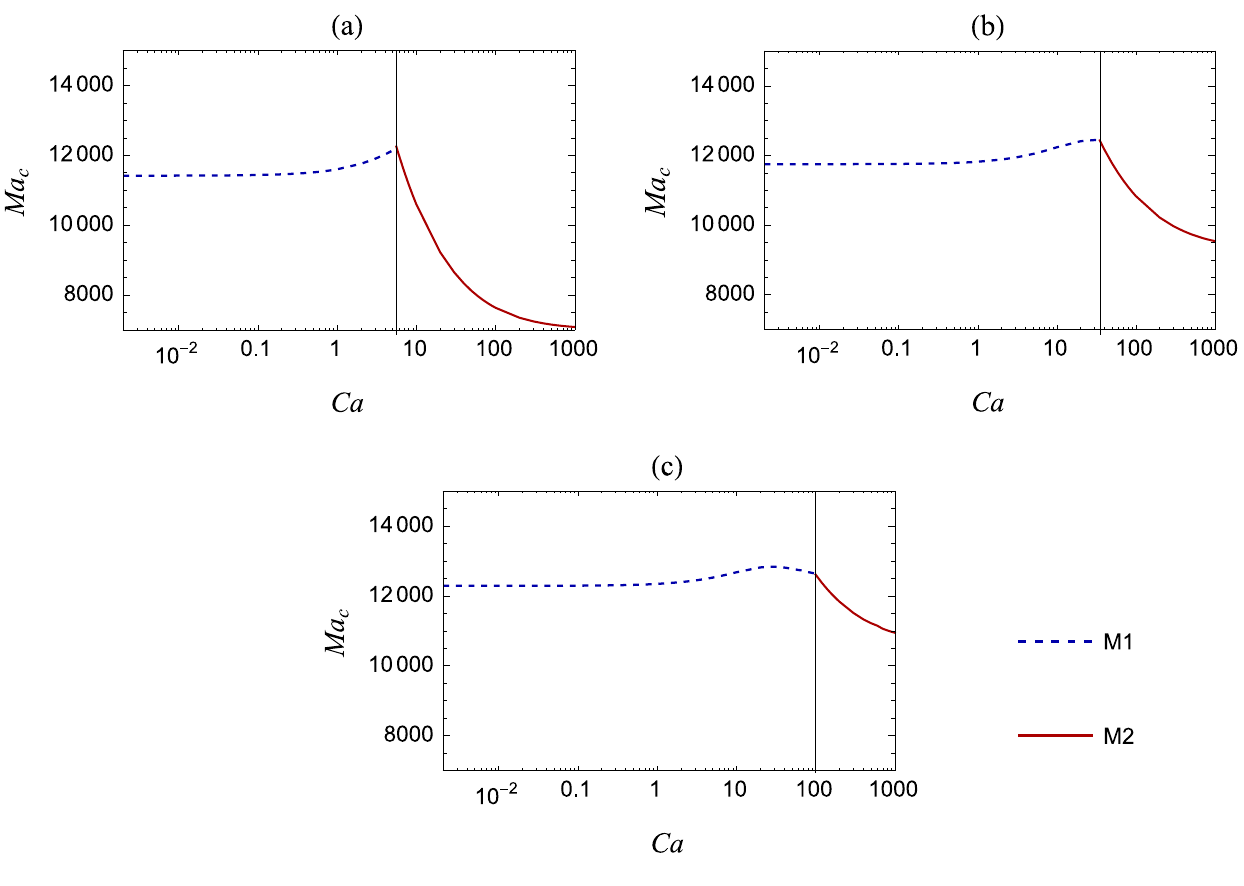}}
  \caption{Effect of $\Pen$ on mode-switching between short wave M1 and M2 modes. (a) $\Pen$ = 1000 (b) $\Pen$ = 2000 (c) $\Pen$ = 2500 Parameter values: $n$ = 1, $m$ = 1.5, $D_r$ = 0.5, $K$ = 0.5, $\Pen$ = 2000, $\gamma$ = 0.5.}
\label{fig:comp_Pe}
\end{figure}

As a final point in this section, we note that the codimension two bifurcation between the short wave modes is affected by other parameter values. As an example, the influence of $\Pen$ ($d_1 U_0/D_1$) is demonstrated in Fig. \ref{fig:comp_Pe}. As $\Pen$ is increased, the switching point shifts to larger values of $Ca$. This implies that increasing the diffusivity of the solute (decreasing $\Pen$) increases the range of criticality of the M2-SW mode. This occurs because diffusion has a stronger stabilizing effect on the M1-SW mode due to its significantly higher wavenumber (Fig. \ref{fig:comp}b). A large wavenumber disturbance has rapid streamwise variations in concentration that are damped out more quickly by diffusion.

\section{Transition from short to long waves}\label{sec:trans}

In \S \ref{sec:longandshort}, it was shown that the M1 branch of eigenvalues gives rise to a long wave instability when $n^2 > m$ and a short wave instability when $n^2 < m$. In this section we examine, via numerical calculations, the transition from short to long waves as $n$ is increased beyond $\sqrt{m}$. We consider two examples, one for a viscosity ratio less than unity and the other for a value larger than unity.
 
\begin{figure}
  \centerline{\includegraphics[scale=1]{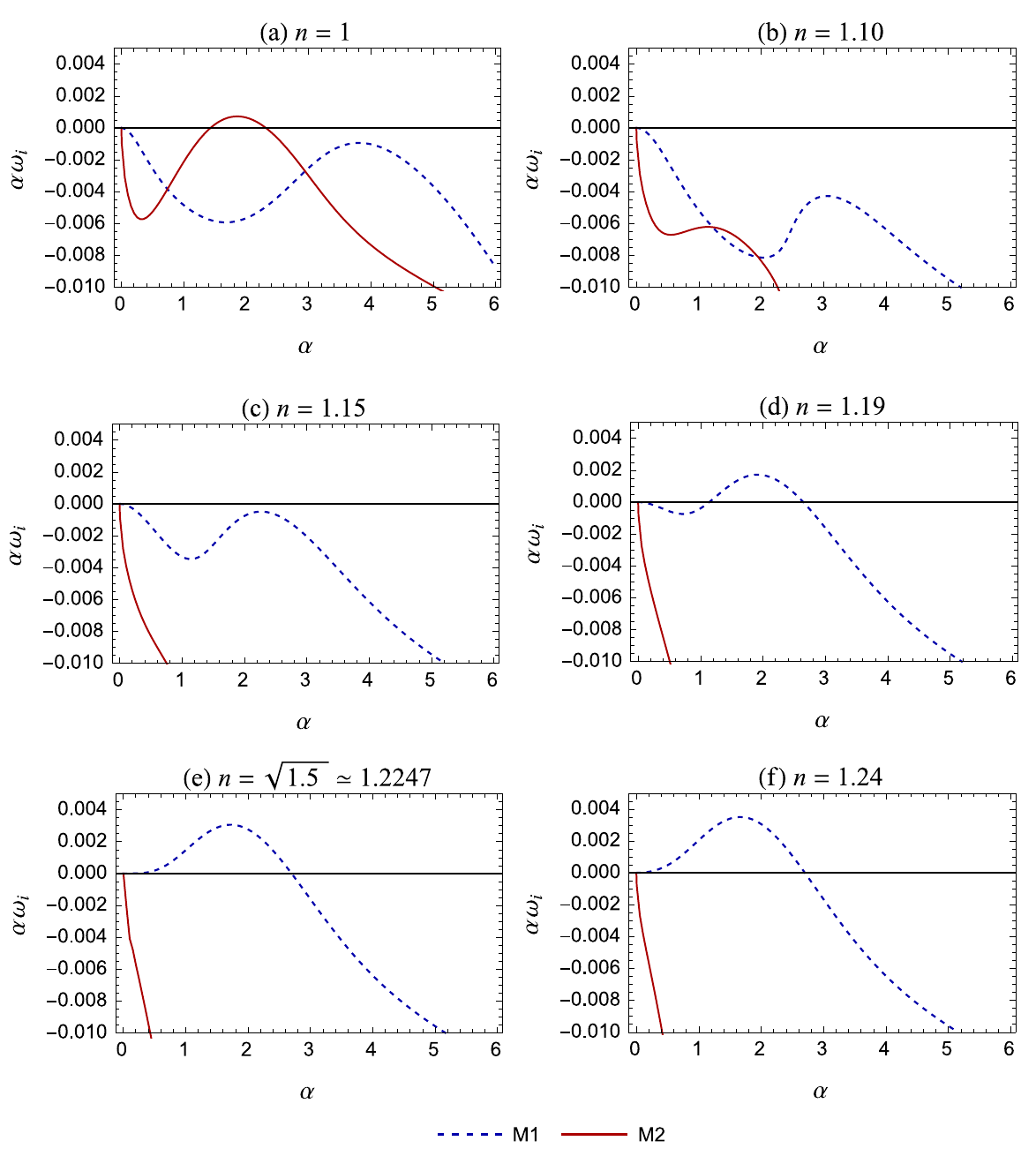}}
  \caption{Transition from a short wave to long wave instability as n is increased beyond $\sqrt{m}$. Parameter values: $m$ = 1.5, $Ca$ = 100, $Ma$ = 11732.65, $D_r$ = 0.5, $K$ = 0.5, $\Pen$ = 2000, $\gamma$ = 0.5.}
\label{fig:transm1p5}
\end{figure}

Fig. \ref{fig:transm1p5} illustrates the transition for a case of $m = 1.5$. The critical value of the thickness ratio is $n=\sqrt{1.5}\approx 1.2247$. Six values of $n$ from 1 to 1.24 are chosen to illustrate the transition. A sufficiently large Capillary number ($Ca=100$) is chosen so that the M2 mode is unstable at the smallest of the selected values of $n$ (Fig. \ref{fig:transm1p5}a). The value of $Ma$ is fixed at 11732.65 for all cases.  

As $n$ increases, with $Ma$ constant, the M2-SW mode is strongly stabilized. The M1-SW mode is also stabilized initially (Fig. \ref{fig:transm1p5}b), but is destabilized again as $n$ is increased further (Fig. \ref{fig:transm1p5}c). Simultaneously, the local maximum growth rate shifts towards longer wavelengths. At $n=1.19$ (Fig. \ref{fig:transm1p5}d) the M1 branch becomes unstable again. This mode is also of the short wave kind, but has significantly larger wavelengths than the M1-SW mode seen at $n=1$(cf. Fig. \ref{fig:transm1p5}a). As $n$ is increased further, the range of unstable wavenumbers of the M1 branch move closer to zero until it becomes a long wave instability (Fig. \ref{fig:transm1p5}f). At the transition ($n=\sqrt{m}$, Fig. \ref{fig:transm1p5}e), the M1 dispersion curve attains zero slope at the origin $\alpha = 0$, in accordance with the asymptotic expression \ref{longexp}.

In \S \ref{sec:longandshort} (table \ref{tab:short_long}), it was shown that the mechanical energy budget of long wave instabilities is dominated by $MAS_I$ while that of the short waves is dominated by $MAS_F$. Hence the transition from short to long waves must be accompanied by a significant change in the energy budget. This is verified in table \ref{tab:trans1}, which presents the mechanical energy-work terms for the local maximum of the M1 dispersion curve at $n = 1, 1.19 \, \rm{and} \, 1.24$. While $MAS_F$ is the dominant positive term for the M1-SW mode ($n = 1$), it decreases as $n$ is increased and becomes negative in the case of long waves ($n=1.24$). Instead, $MAS_I$ becomes the dominant positive term that balances dissipation.

\begin{table}
  \begin{center}
\def~{\hphantom{0}}
  \begin{tabular}{lcccccc}
      $n$  & $\alpha$  & $NOR$ & $TAN_{\mu} $ & \textbf{\textit{$MAS_{I}$}} & \textbf{\textit{$MAS_{F}$}} & $\sum_{j=1}^2E_j^c$ \\[3pt]
       1.00   & 3.805 & $3.09 \times 10^{-7}$ & $-6.74 \times 10^{-3}$ & -0.0213 & 1.0280 & -0.023\\
       1.19   & 1.905 & $-8.70 \times 10^{-5}$ & $5.77 \times 10^{-3}$ & 1.258 & -0.264 & 0.126\\
       1.24   & 1.655 & $-2.80 \times 10^{-4}$ & $3.71 \times 10^{-3}$ & 1.576 & -0.572 & 0.354\\
  \end{tabular}
  \caption{Energy budget for M1 short and long wave modes corresponding to the local maxima of the M1 dispersion curves in Figs. \ref{fig:transm1p5}(a), \ref{fig:transm1p5}(d) and \ref{fig:transm1p5}(f). The mechanical energy terms have been normalized by the magnitude of total viscous dissipation, such that $\sum_{j=1}^2DIS_j=-1$. $KE_i$ and $REY_i$ are identically zero in the creeping flow limit. The stability/instability of the modes can be inferred from the evolution of the concentration energy functional $\sum_{j=1}^2E_j^c$. These terms have been normalized by the total amount of diffusive damping, $\sum_{j=1}^2DIF_j=-1$. Parameter values are given in the caption of Fig. \ref{fig:transm1p5}.}
  \label{tab:trans1}
  \end{center}
\end{table}

Interestingly, table \ref{tab:trans1} shows that the $MAS_I$ term is already larger than the $MAS_F$ term at $n=1.19$, although the M1 dispersion curve still bears the qualitative characteristics of a short wave mode. This mode, which has a relatively longer wavelength than the M1-SW mode at $n=1$, may thus be considered as a distinct intermediate wavelength instability mode. This intermediate mode occurs for a narrow band of $n$ values that separate the M1 short waves from the M1 long waves.

\begin{figure}
  \centerline{\includegraphics[scale=1]{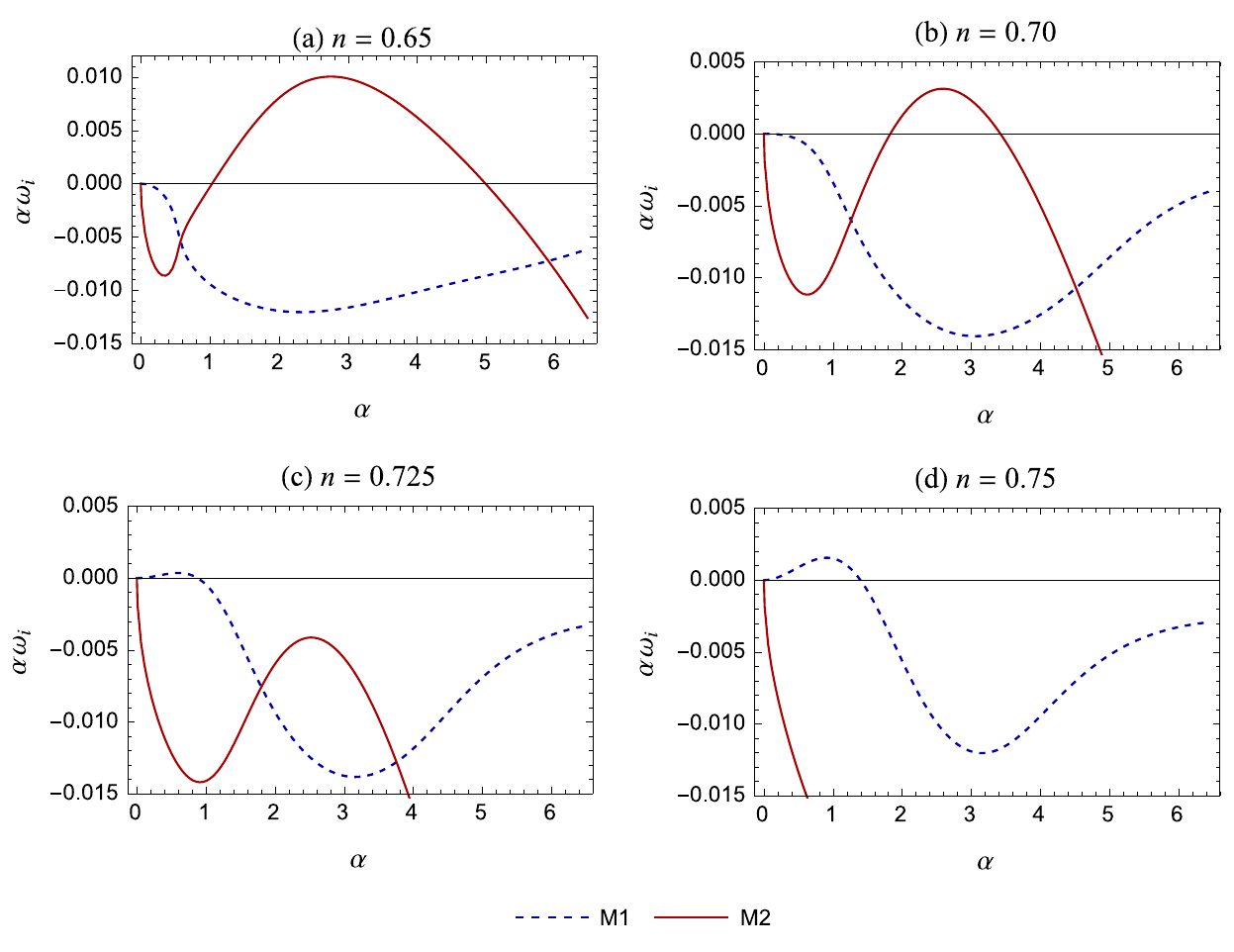}}
  \caption{Transition from a short wave to long wave instability as n is increased beyond $\sqrt{m}$. Parameter values: $m$ = 0.5, $Ca$ = 1000, $Ma$ = 5000, $D_r$ = 1, $K$ = 1, $\Pen$ = 2000, $\gamma$ = 0.}
\label{fig:transmp5}
\end{figure}

An example of the short-long wave transition when $m<1$ is depicted in Fig, \ref{fig:transmp5}. Due to the large value of the Capillary number $Ca=100$, the M2-SW mode is strongly unstable at $n=0.65$. Nevertheless, as $n$ is increased the M2-SW mode is stabilized. The M1 branch of eigenvalues, which has negative growth rates across all $\alpha$ for $n=0.65$, develops a long wave instability when $n$ exceeds $\sqrt{0.5}$ ($=0.707$).

This section has demonstrated how the short wave instability of the M1 branch transitions to a long wave instability, as $n$ is increased beyond $\sqrt{m}$. The M2-SW mode is generally stabilized as $n$ is increased, provided $Ma$ is constant. If $Ma$ is also increased, however, then the M2-SW can remain unstable despite the stabilizing influence of increasing $n$. In such a case, the system will be unstable to long and short waves simultaneously (as shown in Fig. \ref{fig:long}b).

\section{Necessity of a base state transverse concentration gradient}\label{sec:flux}

In this section, we investigate whether the mere presence of soluble surfactant at the interface is sufficient to cause a Marangoni instability, or whether a finite transverse concentration gradient is also required. The base state concentration field has a finite gradient if the plates are maintained at non-equilibrium concentrations ($\gamma \ne 1/K$). The long wave analysis, in \S \ref{sec:longwave}, demonstrated that a finite concentration gradient is necessary for long wave Marangoni modes to be unstable. Here we check whether this remains a prerequisite for short wave Marangoni instability modes as well.

\begin{figure}
  \centerline{\includegraphics[scale=1]{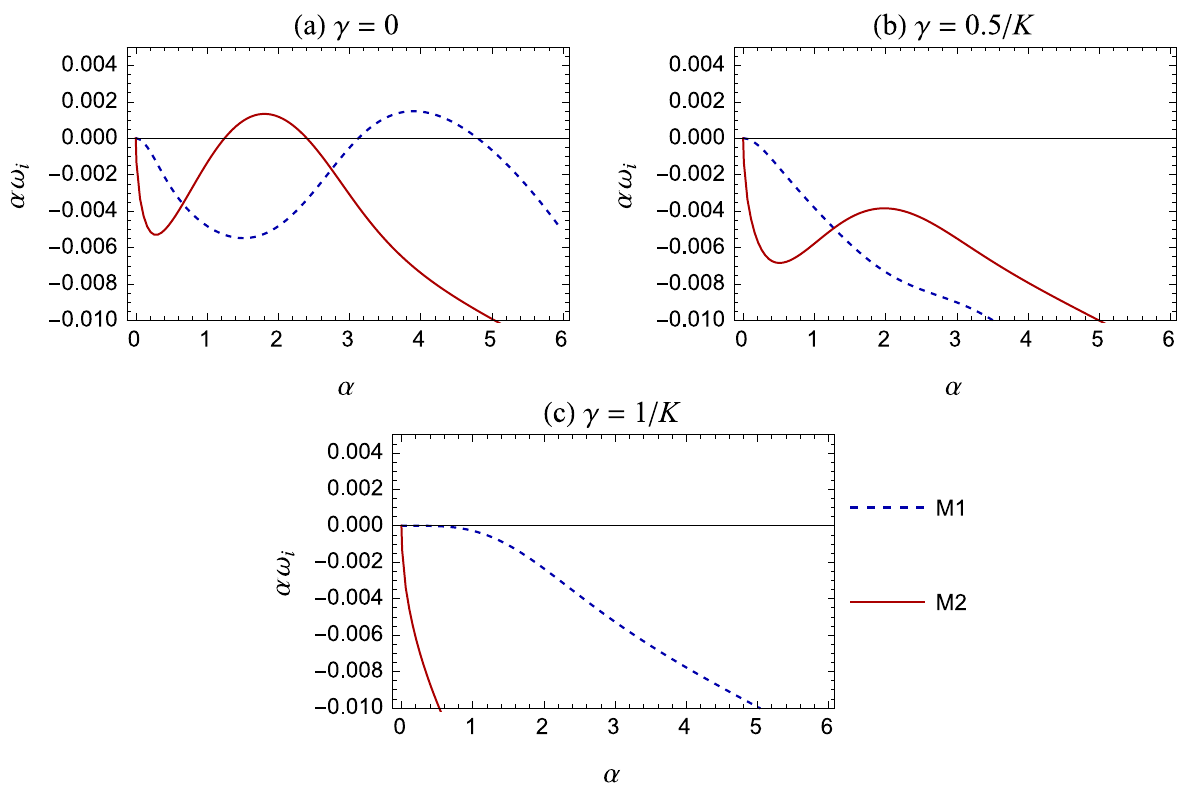}}
  \caption{Effect of the inter-fluid flux on the short wave instability. Both M1 and M2 short waves are stabilized as the difference of the concentration at the walls from equilibrium is decreased. $\gamma$ = 1/K corresponds to a base state without inter-fluid flux, as the fluids are in equilibrium in this case and have non-varying concentration profiles. Parameter values: $m$ = 1.5, $n$ = 1, $Ca$ = 100, $Ma$ = 10000, $D_r$ = 0.5, $K$ = 1.2, $\Pen$ = 2000.}
\label{fig:flux_short}
\end{figure}

\begin{figure}
  \centerline{\includegraphics[scale=1]{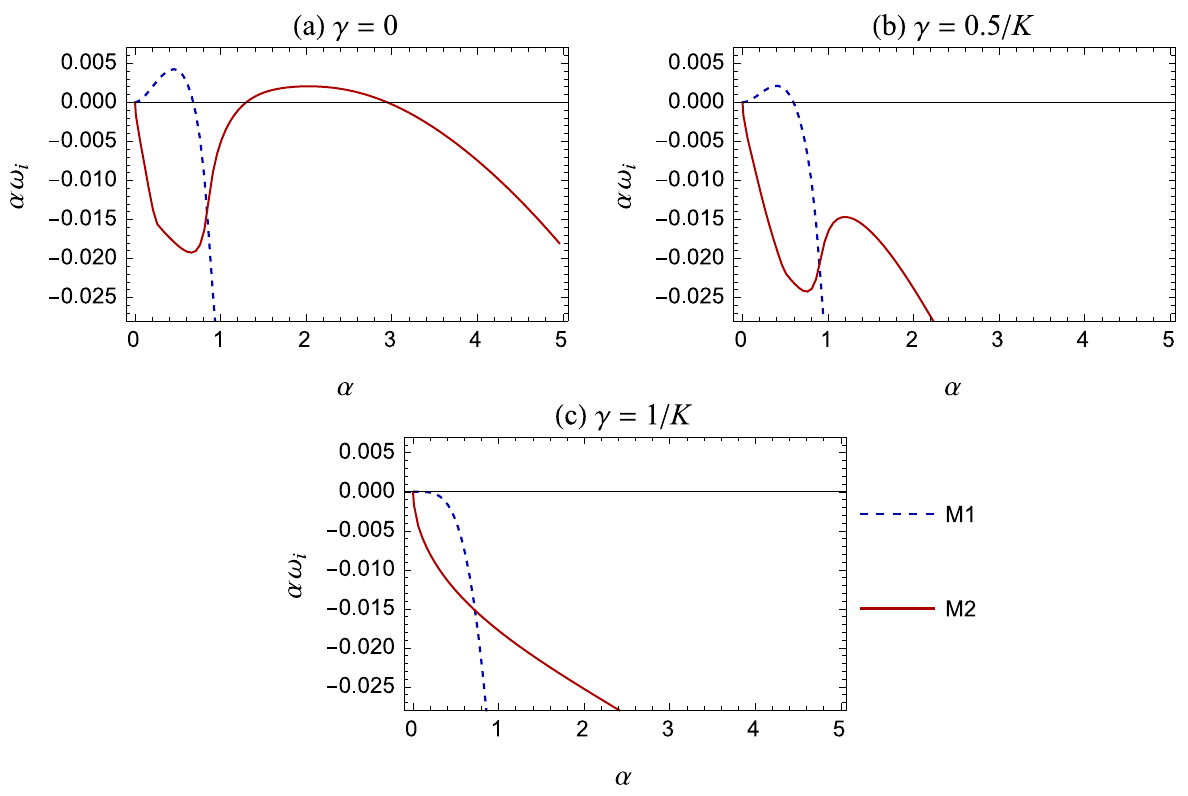}}
  \caption{Effect of the inter-fluid flux on the long wave instability. The long wave M1 mode, as well as the M2 short wave mode, is stabilized as the difference of the concentration at the walls from equilibrium is decreased. $\gamma$ = 1/K corresponds to a base state without inter-fluid flux, as the fluids are in equilibrium in this case and have non-varying concentration profiles. Parameter values: $m$ = 0.9, $n$ = 1.3, $Ca$ = 1, $Ma$ = 10000, $D_r$ = 0.5, $K$ = 1.2, $\Pen$ = 1000.}
\label{fig:flux_long}
\end{figure}

Figs. \ref{fig:flux_short} and \ref{fig:flux_long} demonstrate the stabilizing effect of decreasing the base concentration gradient. Each figure considers three cases, corresponding to $\gamma=0,\,0.5/K,$ and $1/K$ (equilibrium). The parameter values in Fig. \ref{fig:flux_short} are selected so that M1-SW and M2-SW modes are unstable in the presence of base state mass transfer ($\gamma=0$). Fig. \ref{fig:flux_long} corresponds to the case of unstable M1-LW and M2-SW. In both figures, the system is seen to become stable when the plates are maintained at equilibrium concentrations ($\gamma=1/K$).

The necessity of a transverse base state concentration gradient for solutal Marangoni instability can be understood with the aid of the energy balance equations (cf. \S \ref{sec:energy}).  In the mechanical energy balance \eqref{mechenergy}, the terms contributing to the Marangoni instability are $MAS_F$ and $MAS_I$. When the base state concentration gradient is zero $(d\overline{c}_j/dy=0)$, interface deformation does not generate concentration variations along the interface. Thus $MAS_I$ is identically zero. $MAS_F$ accounts for concentration perturbations $\widehat{c_i}$ that are coupled to the disturbance flow within the fluids. In the absence of a base state concentration gradient, convection by the disturbance flow is negligible in comparison with the stabilizing effects of diffusion and viscous dissipation. This is reflected by the vanishing of $CONT_j$ - the term that represents the contribution of convective effects to the growth of the concentration energy functional (cf. \eqref{concenergy}). Thus both mechanisms for generating concentration variations along the interface are ruled out. This explains why a finite concentration gradient must exist across the phases for the solutal Marangoni instability to occur.

\section{Influence of finite inertia}\label{sec:inertia}

Thus far we have focused on the creeping flow regime, wherein we have identified three distinct instability modes - two short waves and one long wave. In this section, we briefly consider the  case of small but finite $\Rey$. A thorough analysis of the effects of inertia would merit a separate study. Here we discuss only a few specific cases, which illustrate the interplay of the viscosity-induced mode and the solutal Marangoni modes. 

We pursue two different lines of inquiry. Firstly, we investigate the influence of inertia on the three Marangoni instability modes. Starting with an unstable creeping flow, the value of $\Rey$ is increased and its effect on the three modes is observed.  Secondly, we investigate the effect of introducing soluble surfactant into a system which is already unstable to the viscosity-induced mode. These two cases are considered separately in the following subsections.

\subsection{Effect of inertia on the solutal Marangoni instability}

In this subsection, the effect of inertia on the three solutal Marangoni modes is investigated for two examples: one for $n^2<m$ and the other for $n^2>m$. 

\begin{figure}
  \centerline{\includegraphics[scale=1]{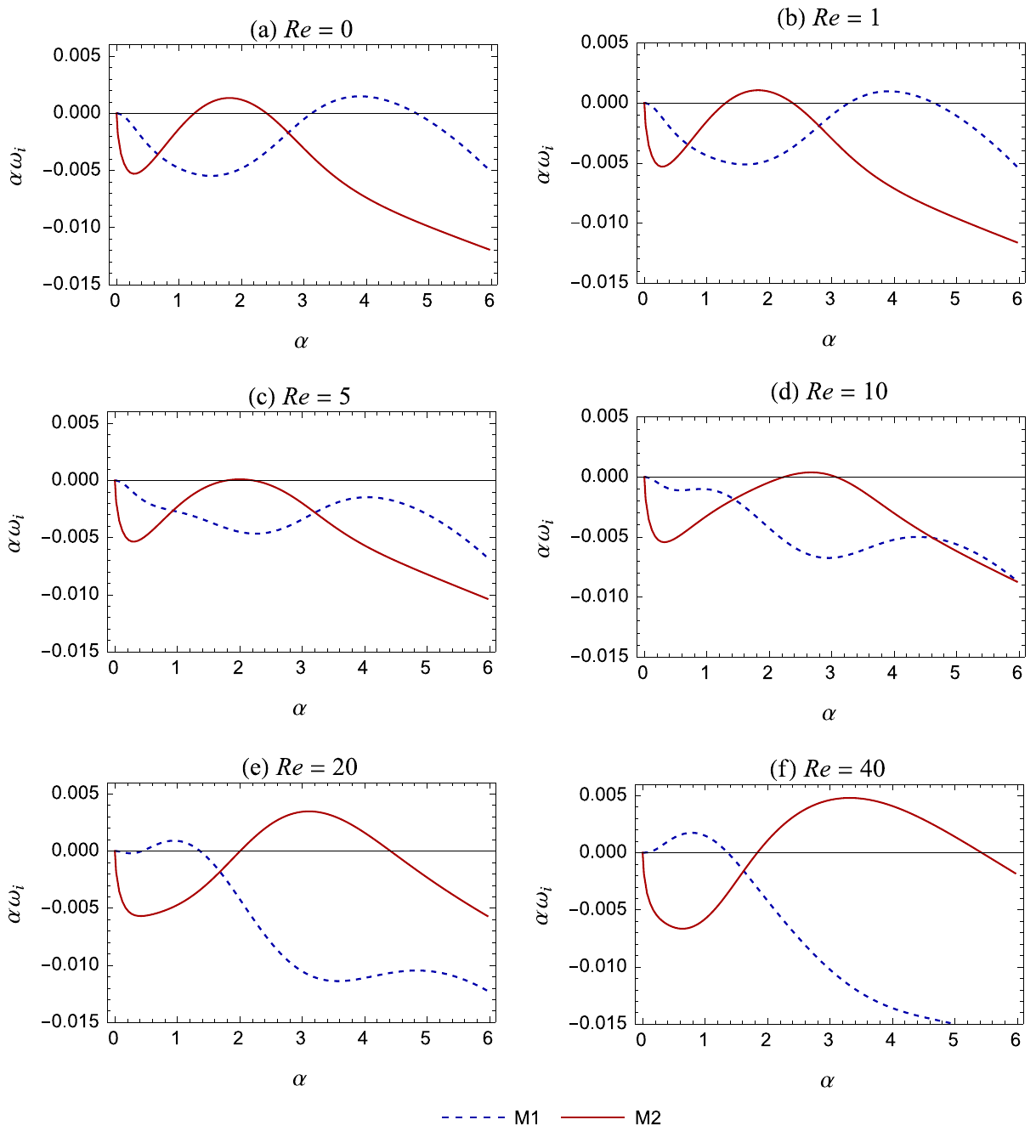}}
  \caption{Effect of inertia on the solutal Marangoni instability when the M1 mode is short wave. Parameter values: $Ma$ = 10000, $m$ = 1.5, $n$ = 1, $Ca$ = 100, $D_r$ = 0.5, $K$ = 1.2, $\gamma = 0$, $\Pen$ = 2000.}
\label{fig:inertia_Re_short}
\end{figure}

The first case of $n^2<m$ is depicted in Fig. \ref{fig:inertia_Re_short}). Here, creeping flow is unstable to short wave M1-SW and M2-SW modes ($n^2<m$). In Figs. \ref{fig:inertia_Re_short}(a)-(f) $\Rey$ is increased sequentially from 0 to 40. The results for $\Rey=1$ (Fig. \ref{fig:inertia_Re_short}(b)) are similar to the creeping flow case (Fig. \ref{fig:inertia_Re_short}(a)). However, increasing $\Rey$ to 5 (Fig. \ref{fig:inertia_Re_short}(c)) significantly stabilizes both modes. The effect of further increasing $\Rey$ differs for the two modes. The M2-SW mode becomes unstable again and its growth rate increases with $\Rey$. On the other hand, the M1-SW mode undergoes a transition to long waves as $\Rey$ is increased beyond 10 ((Figs. \ref{fig:inertia_Re_short}(d))-(f)). This behaviour is in accordance with the long wave asymptotic prediction \eqref{longexp} that inertia destabilizes long waves when $m>1$ and $n^2<m$.

Mechanical energy budget (cf. \eqref{mechenergy}) calculations for the local maximum of the M1 and M2 dispersion curves, corresponding to Figs. \ref{fig:inertia_Re_short}(a) and \ref{fig:inertia_Re_short}(f) is presented in table \ref{tab:inertia_short}. The dominant work term of the M1 mode changes from $MAS_F$ to $MAS_I$ in accordance with the transition from short waves to long waves. The energy budgets of the most unstable M2 mode for $\Rey=0$ and $\Rey=40$ are qualitatively the same ($MAS_F$ is the dominant positive term). This confirms that the M2 short wave mode that is unstable at finite $\Rey$ is qualitatively the same as the M2-SW mode observed in creeping flow.

\begin{table}
  \begin{center}
\def~{\hphantom{0}}
  \begin{tabular}{lccccccccc}
      Mode & $\Rey$ & $\alpha$ & $KE_1+KE_2$ & $REY_1+REY_2$ & $NOR$ & $TAN_{\mu} $ & \textbf{\textit{$MAS_{I}$}} & \textbf{\textit{$MAS_{F}$}} \\[3pt]
      M1 & 0 & 3.92 & 0 & 0 & $-2.4 \times 10^{-7}$ & $-4.62 \times 10^{-3}$ & -0.011 & 1.016 \\
      M1 & 40 & 0.77 & 0.002 & -0.029 & $-7.45 \times 10^{-5}$ & $-6.99 \times 10^{-2}$ & 0.730 & 0.371 \\
      M2 & 0 & 1.82 & 0 & 0 & $-3.29 \times 10^{-5}$ & 0.115 & -0.108 & 0.994 \\
      M2 & 40 & 3.32 & 0.002 & -0.024 & $-2.15 \times 10^{-4}$ & 0.108 & 0.090 & 0.828 \\
  \end{tabular}
  \caption{Energy budget for M1 and M2 modes corresponding to the local maxima of the dispersion curves in Figs. \ref{fig:inertia_Re_short}(a) and \ref{fig:inertia_Re_short}(f). The values have been normalized by the magnitude of total dissipation so that $(DIS{_1}+ DIS{_2})=-1$. Parameter values are given in the caption of Fig. \ref{fig:inertia_Re_short}.}
  \label{tab:inertia_short}
  \end{center}
\end{table}

\begin{figure}
  \centerline{\includegraphics[scale=1]{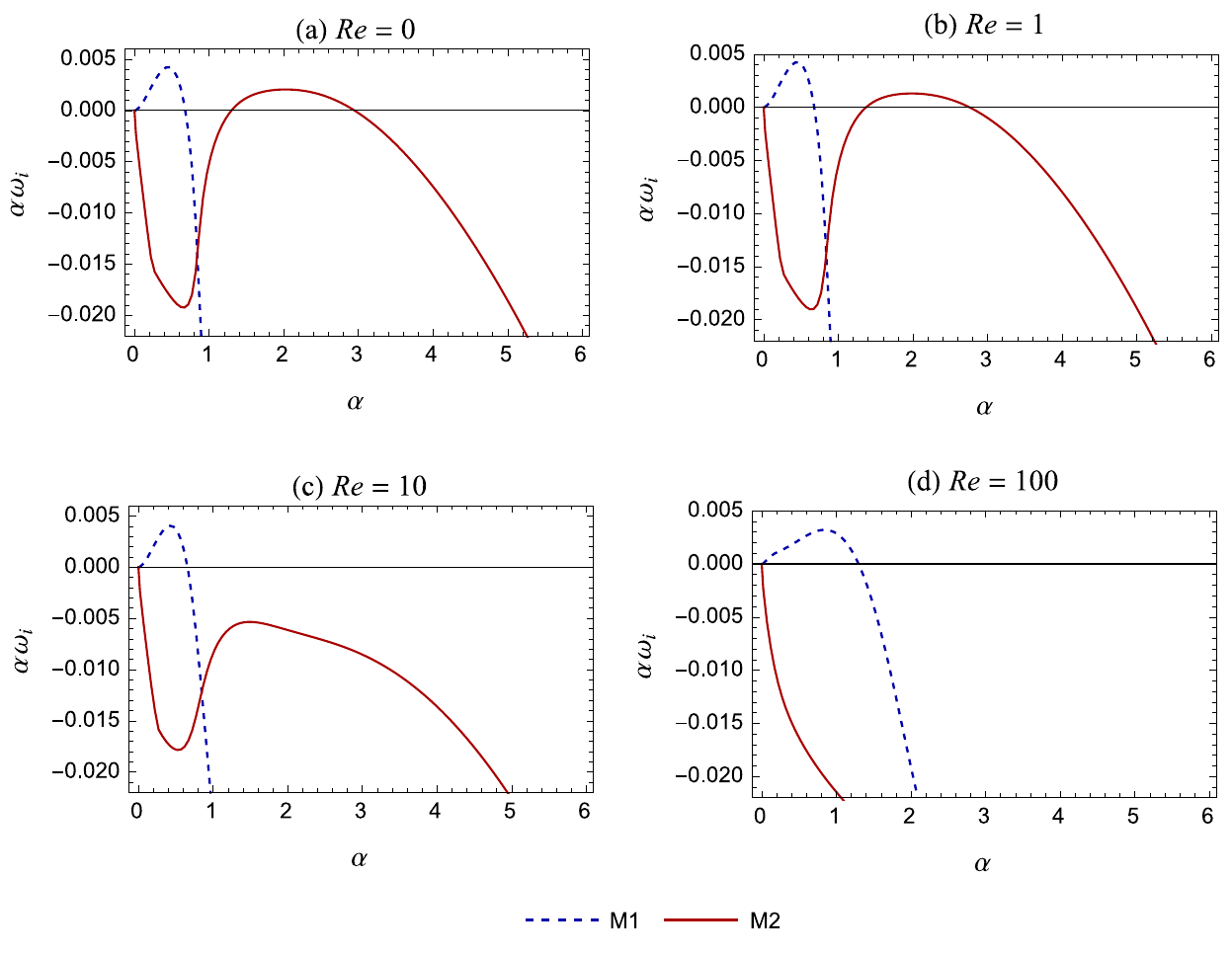}}
  \caption{Effect of inertia on the solutal Marangoni instability when the M1 mode is long wave. Parameter values: $Ma$ = 10000, $m$ = 0.9, $n$ = 1.3, $Ca$ = 1, $D_r$ = 0.5, $K$ = 1.2, $\gamma = 0$, $\Pen$ = 1000.}
\label{fig:inertia_Re_long}
\end{figure}

Fig. \ref{fig:inertia_Re_long} demonstrates the effect of inertia for a case of $n^2>m$, wherein the long wave M1 mode and the M2-SW mode are unstable in the creeping flow limit (Fig. \ref{fig:inertia_Re_long}(a)). As $\Rey$ is increased, the M2-SW mode is completely stabilized. The long wave M1 instability, on the other hand, remains unstable. In fact, the growth rate is greater at $\Rey=1$ than at creeping flow. This is in accordance with the prediction of the long wave expansion \eqref{longexp} for the parameter values of Fig. \ref{fig:inertia_Re_long} ($n^2>m$ and $m<1$, cf. the figure caption). The effect of further increase in $\Rey$ cannot be predicted by the asymptotic expansion, since it was derived for the case of $\Rey$ being $O(1)$. Fig. \ref{fig:inertia_Re_long}(c)-(d) show that long wave M1 modes continue to remain unstable, as $\Rey$ is increased to 100, without significant change in their growth rates. Even at $\Rey=100$, these long wave modes are primarily driven by Marangoni stresses associated with interface deformation ($MAS_I$), as shown by the energy budget calculations presented in table \ref{tab:inertia_long}.

These two examples demonstrate that inertia affects the M1 and M2 Marangoni modes quite differently. Increasing $\Rey$ can qualitatively modify the M1 branch of eigenvalues and cause a transition between M1 short waves and M1 long waves. This strong influence over the M1 branch is due to the fact that the viscosity-induced mode manifests itself through the M1 branch of eigenvalues. (When $Ma=0$ and $\Rey \ne 0$, the unstable viscosity induced mode emerges as long waves on the M1 branch, as shown in \S \ref{sec:inertiaMa}.) On the other hand, increasing $\Rey$ does not modify the basic qualitative features of the M2-SW mode. The effect on its growth rate, however, can be non-monotonic, causing intermediate stabilization of the system over a narrow range of $\Rey$ as demonstrated by Fig. \ref{fig:inertia_Re_short}(b)-(d).

\begin{table}
  \begin{center}
\def~{\hphantom{0}}
  \begin{tabular}{lccccccccc}
      Mode & $\Rey$ & $\alpha$ & $KE_1+KE_2$ & $REY_1+REY_2$ & $NOR$ & $TAN_{\mu} $ & \textbf{\textit{$MAS_{I}$}} & \textbf{\textit{$MAS_{F}$}} \\[3pt]
      M1 & 0 & 0.47 & 0 & 0 & -0.014 & -0.033 & 1.676 & -0.629 \\
      M1 & 100 & 0.82 & 0.009 & -0.029 & -0.017 & -0.037 & 1.552 & -0.459 \\
  \end{tabular}
  \caption{Energy budget for long wave M1 modes corresponding to the local maxima of the dispersion curves in Figs. \ref{fig:inertia_Re_long}(a) and \ref{fig:inertia_Re_long}(d). The values have been normalized by the magnitude of total dissipation so that $(DIS{_1}+ DIS{_2})=-1$. Parameter values are given in the caption of Fig. \ref{fig:inertia_Re_long}.}
  \label{tab:inertia_long}
  \end{center}
\end{table}

\subsection{Effect of soluble surfactant on the viscosity-induced instability at finite $Re$}\label{sec:inertiaMa}

In the absence of soluble surfactant effects ($Ma=0$), a finite $Re$ flow is unstable to the viscosity-induced instability, provided $m>1$ ($m<1$) when $n^2<m$ ($n^2>m$) (cf. \S \ref{sec:longwave}). In this section, we follow the changes in stability characteristics which occur on applying a transverse gradient of soluble surfactant. Two examples are studied in this section, corresponding to $n^2<m$ (with $m>1$) and $n^2>m$ (with $m<1$). Based on the preceding creeping flow analysis, the M1-SW Marangoni mode is expected to play a more significant role than the M1-LW mode in the first case, while the opposite is true of the second case (cf. \S \ref{sec:longandshort}). The M2-SW mode could impact the flow in either case.  

\begin{figure}
  \centerline{\includegraphics[scale=1]{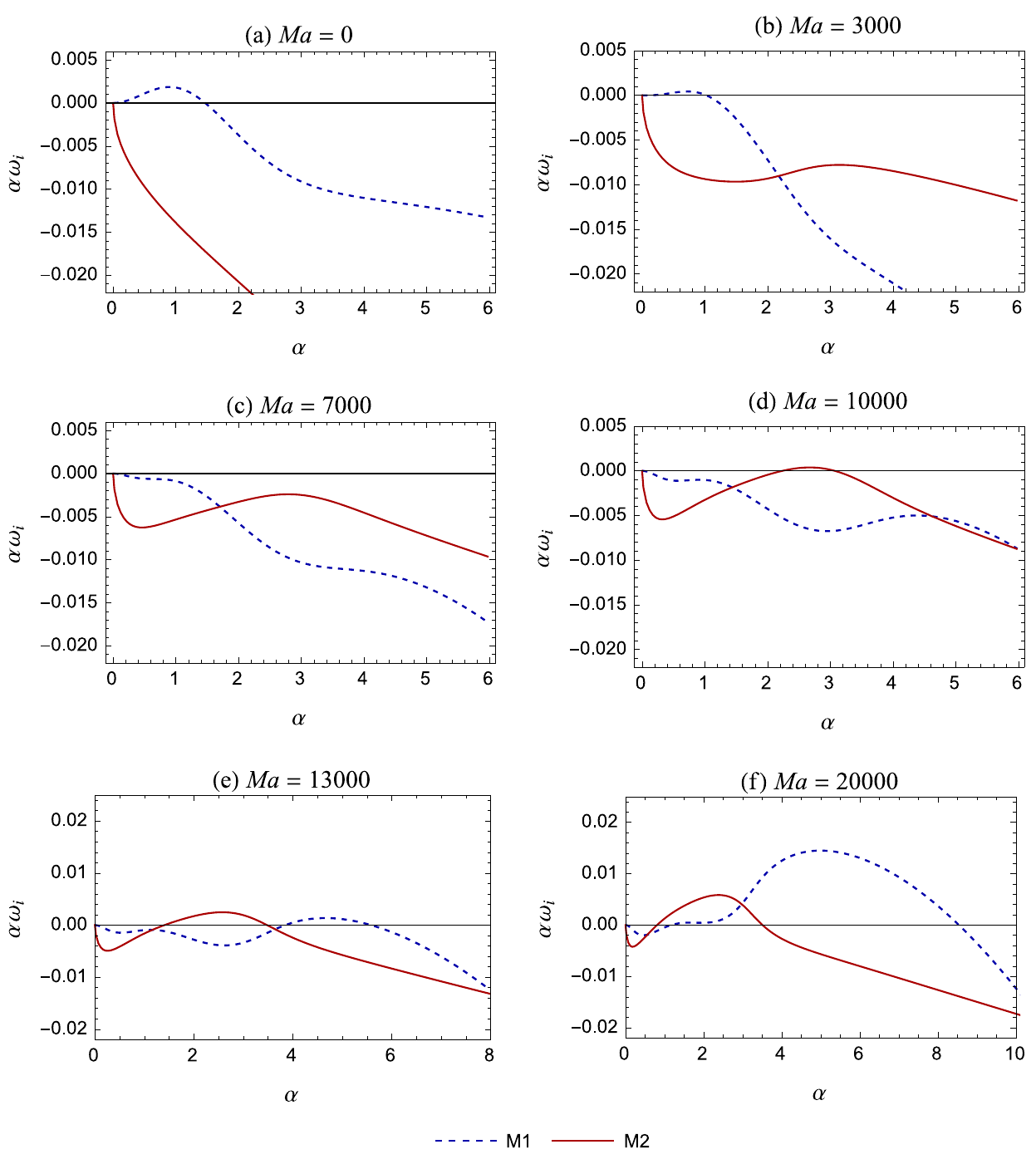}}
  \caption{Effect of soluble surfactant on the viscosity-induced interfacial instability. The chosen parameter values correspond to the case of $n^2<m$, for which the short wave M1-SW and M2-SW modes are unstable in the creeping flow limit. Parameter values: $\Rey$ = 10, $m$ = 1.5, $n$ = 1, $Ca$ = 100, $D_r$ = 0.5, $K$ = 1.2, $\gamma = 0$, $\Pen$ = 2000.}
\label{fig:inertia_Ma_short}
\end{figure}

Fig. \ref{fig:inertia_Ma_short} corresponds to the case of $n^2<m$. When $Ma=0$ (Fig. \ref{fig:inertia_Ma_short}(a)), the long wave viscosity-induced mode is unstable. The identity of this mode is confirmed by energy budget calculations, presented in table \ref{tab:inertia_Ma_short}, which show that the dominant work term is $TAN_{\mu}$ (cf. \citet{boomkamp2}). Since the viscosity-induced mode manifests itself via the M1 branch of eigenvalues, the M1 Marangoni modes (M1-SW in this case) are expected to have a direct impact on it. Indeed, on increasing $Ma$, the viscosity induced mode is suppressed and undergoes a transition to the short wave M1 Marangoni mode (Fig. \ref{fig:inertia_Ma_short}(b-e)). This transition is confirmed by the energy budget calculations presented in table \ref{tab:inertia_Ma_short}. At $Ma=13000$, $TAN_{\mu}$ is negligible in comparison with $MAS_F$, which is the dominant positive term. This is characteristic of the energy budget of the short wave solutal Marangoni mode (\S \ref{sec:comp}). 

Along with the emergence of the M1-SW mode, the M2-SW mode also becomes unstable as $Ma$ is increased beyond 10000 (Fig. \ref{fig:inertia_Re_long}(d)-(f)). Therefore, as $Ma$ is increased, solutal Marangoni effects dominate and become the primary cause for instability of the flow.

\begin{table}
  \begin{center}
\def~{\hphantom{0}}
  \begin{tabular}{lccccccccc}
      Mode & $Ma$ & $\alpha$ & $KE_1+KE_2$ & $REY_1+REY_2$ & $NOR$ & $TAN_{\mu} $ & \textbf{\textit{$MAS_{I}$}} & \textbf{\textit{$MAS_{F}$}} \\[3pt]
      M1 & 0 & 0.92 & $1.51 \times 10^{-3}$ & $-7.51 \times 10^{-3}$ & $-1.63 \times 10^{-3}$ & 1.010 & 0 & 0 \\
      M1 & 13000 & 4.67 & $1.17 \times 10^{-4}$ & $-1.47 \times 10^{-3}$ & $-2.91 \times 10^{-6}$ & $1.19 \times 10^{-3}$ & -0.182 & 1.182 \\
  \end{tabular}
  \caption{Energy budget showing the transition from the viscosity induced mode to the solutal Marangoni M1-SW mode. Calculations correspond to the fastest growing mode in Figs. \ref{fig:inertia_Ma_short}(a) and \ref{fig:inertia_Ma_short}(e). The values have been normalized by the magnitude of total dissipation so that $(DIS{_1}+ DIS{_2})=-1$. Parameter values are given in the caption of Fig. \ref{fig:inertia_Ma_short}.}
  \label{tab:inertia_Ma_short}
  \end{center}
\end{table}

\begin{figure}
  \centerline{\includegraphics[scale=1]{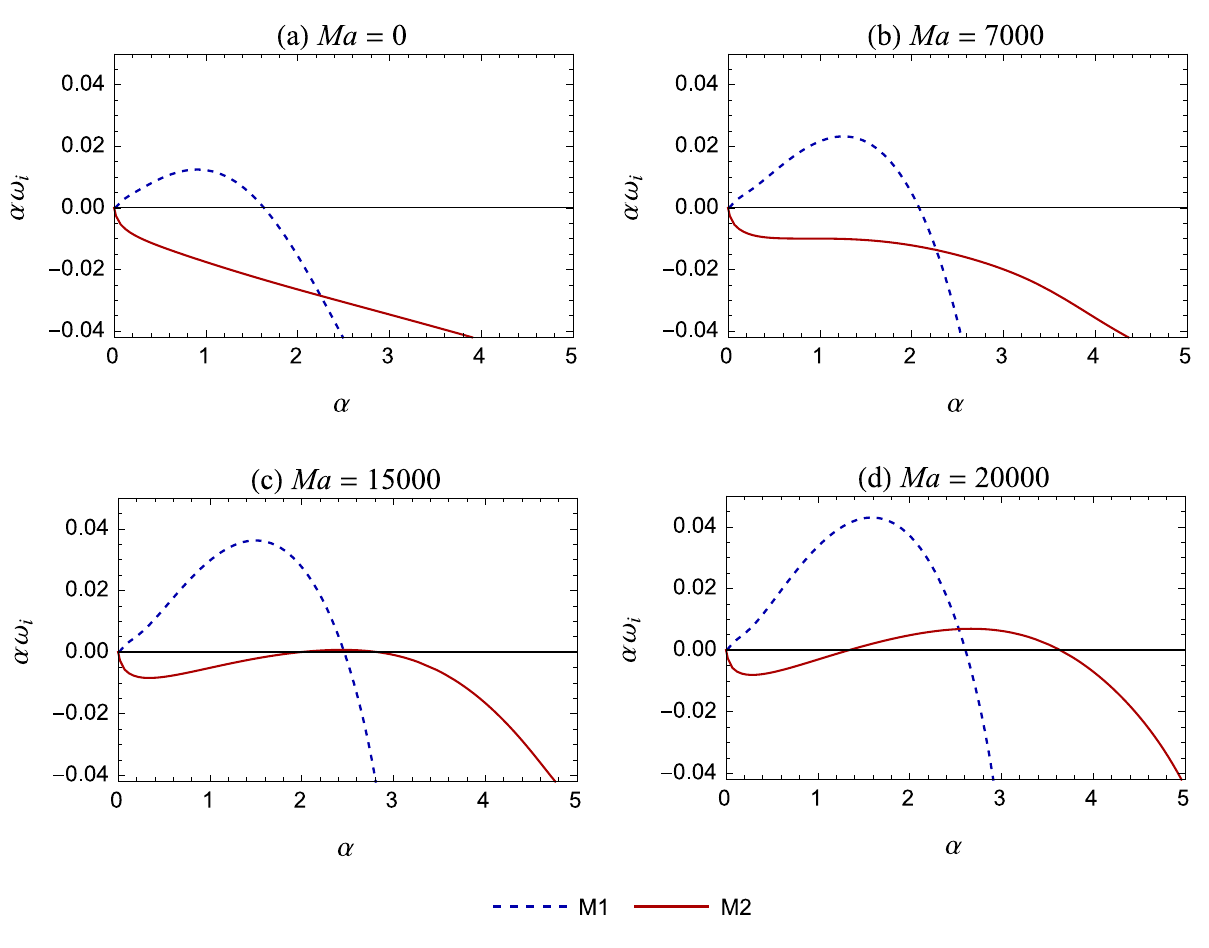}}
  \caption{Effect of soluble surfactant on the viscosity-induced interfacial instability. The chosen parameter values correspond to the case of $n^2>m$, for which the long wave M1 and short wave M2-SW instabilities are prevalent in the creeping flow limit. Parameter values: $\Rey$ = 100, $m$ = 0.5, $n$ = 1.3, $Ca$ = 1, $D_r$ = 0.5, $K$ = 1.2, $\gamma = 0$, $\Pen$ = 1000.}
\label{fig:inertia_Ma_long}
\end{figure}

\begin{figure}
  \centerline{\includegraphics[scale=1]{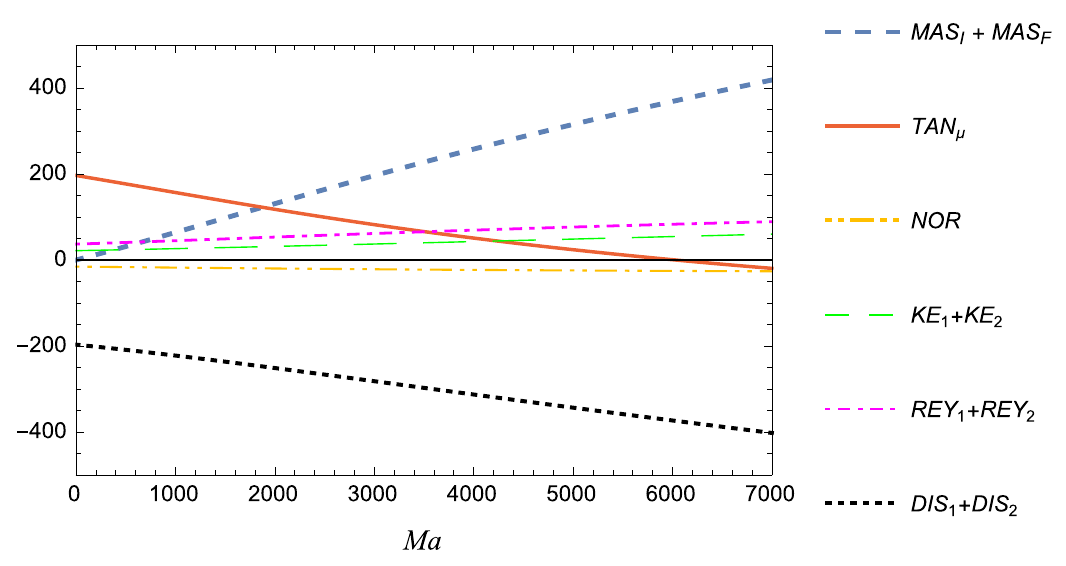}}
  \caption{Variation of the energy budget as $Ma$ is varied, for $\alpha$ = 1 in Fig. \ref{fig:inertia_Ma_long}. Parameter values are given in the caption of Fig. \ref{fig:inertia_Ma_long}.}
\label{fig:energy_Ma}
\end{figure}

An example in which $n^2>m$ is considered in Fig. \ref{fig:inertia_Ma_long}. As in the previous case (Fig. \ref{fig:inertia_Ma_short}), short waves from the M2 branch (M2-SW mode) become unstable on increasing $Ma$. The growth rates of the long waves increase with $Ma$, but the M1 dispersion curve does not change qualitatively. To investigate the nature of this long wave instability, the variation of the energy budget with $Ma$ for $\alpha=1$ is plotted in Fig. \ref{fig:energy_Ma}. For small $Ma$, $TAN_{\mu}$ is dominant, indicating that the instability is due to the viscosity induced mode. As $Ma$ is increased, $TAN_{\mu}$ decreases while the Marangoni stress terms $MAS_I+MAS_F$ increase. The crossover point occurs around $Ma=2000$. By $Ma=7000$ (Fig. \ref{fig:inertia_Ma_long}(b)), $TAN_{\mu}$ is negative and solutal Marangoni forces are the dominant cause of instability. The energy budget of the fastest growing long wave modes for $Ma=0$ and $Ma=20000$ (Figs. \ref{fig:inertia_Ma_long}(a) and \ref{fig:inertia_Ma_long}(d) respectively) are presented in table \ref{tab:inertia_Ma_long}. While $TAN_{\mu}$ is the largest positive term for $Ma=0$, the energy budget for $Ma=20000$ is dominated by the $MAS_I$ term, which is characteristic of the long wave solutal Marangoni instability M1-LW (\S \ref{sec:longandshort}).

\begin{table}
  \begin{center}
\def~{\hphantom{0}}
  \begin{tabular}{lccccccccc}
      Mode & $Ma$ & $\alpha$ & $KE_1+KE_2$ & $REY_1+REY_2$ & $NOR$ & $TAN_{\mu} $ & \textbf{\textit{$MAS_{I}$}} & \textbf{\textit{$MAS_{F}$}} \\[3pt]
      M1 & 0 & 0.92 & 0.119 & 0.183 & -0.069 & 1.005 & 0 & 0 \\
      M1 & 20000 & 1.57 & 0.168 & 0.165 & -0.086 & -0.281 & 1.254 & 0.117 \\
  \end{tabular}
  \caption{Energy budget showing the transition from the viscosity induced mode to the long wave solutal Marangoni M1 mode. Calculations correspond to the fastest growing mode in Figs. \ref{fig:inertia_Ma_long}(a) and \ref{fig:inertia_Ma_long}(d). The values have been normalized by the magnitude of total dissipation so that $(DIS{_1}+ DIS{_2})=-1$. Parameter values are given in the caption of Fig. \ref{fig:inertia_Ma_long}.}
  \label{tab:inertia_Ma_long}
  \end{center}
\end{table}

In summary, introducing a gradient of soluble surfactant into an unstable small $\Rey$ flow causes a transition from the viscosity induced mode to the solutal Marangoni instability (provided $Ma$ is sufficiently large). When the M1-SW Marangoni mode is dominant ($n^2<m$), this transition may cause the system to become stable in an intermediate range of $Ma$ (cf. Fig. \ref{fig:inertia_Ma_short}). This is because the long wave viscosity induced mode is first suppressed before arising again as the short wave M1-SW Marangoni mode. On the other hand, when the long wave M1-LW Marangoni mode is dominant ($n^2>m$) the flow becomes increasingly unstable to long wave disturbances, on increasing $Ma$.  The viscosity-induced mode eventually transitions to the Marangoni M1-LW mode.

\section{Comparison with previous work}\label{sec:compare}

In this section, we compare our results with the closely related study by \citet{Wei2006}, on the thermocapillary instability of Couette flow. We also discuss the implications of neglecting concentration perturbations caused by interface deformation, as is done in \citet{You2014a}.

\citet{Wei2006} has studied the stability of two-phase layered \textit{Couette} flow between flat plates, which are maintained at different temperatures. The temperature variation across the fluids in the base state is analogous to the variation of solute concentration in the present solutal Marangoni problem. In fact, our entire analysis is valid for thermocapillary instabilities in \textit{Poiseuille} flow, provided we set $K=1$ (as temperature is continuous at the interface) and restrict $Ma$ to positive values (since interfacial tension decreases with temperature). Also $D_r$ must be identified with the ratio of thermal diffusivities. 

\citet{Wei2006} considered the problem in the ``thin layer limit", in which one fluid layer is very thin in comparison with the other fluid layer. He obtained a long wave instability mode, which is driven by thermocapillary stresses caused by deformation of the interface in the presence of a base state temperature gradient. This is the thermocapillary analogue of the M1-LW mode (\S \ref{sec:longandshort}). The mode was found to be unstable when the thin layer is heated. If we consider fluid one to be the thin fluid, then we have $n\gg 1$ and consequently $n^2>m$ for any finite viscosity ratio $m$. Under these conditions our analysis  predicts instability of the long wave M1-LW mode (cf. \eqref{longexp}) if the concentration at plate 1 is greater than that at plate 2 ($\gamma<1$), which is analogous to heating the thin fluid. 

\citet{Wei2006} also finds that the growth rate for small wavenumber is linearly proportional to $D_r$. Our long wavelength asymptotic result \eqref{longexp} shows that the growth rate varies as $n^2 D_r/(D_r+n)^2$, which implies linear proportionality with $D_r$ in the thin layer limit of large $n$. (Note that the wavenumber should be rescaled with the depth of fluid 2 before taking the limit of large $n$ in \eqref{longexp}.) Furthermore, at small nonzero $\Rey$, it was observed that the viscosity induced mode reinforces the long wave thermocapillary mode if the thin layer is more viscous ($m<1$), and suppresses it if the thin layer is less viscous ($m>1$). This interaction is entirely analogous to that predicted by \eqref{longexp} for the case of a heated thin layer ($n\gg1$ and $\gamma<1$). Thus our results for the M1-LW mode are in qualitative agreement with \citet{Wei2006}. Further quantitative comparison is not possible since \citet{Wei2006} studied Couette flow while we have analysed Poiseuille flow.

\begin{figure}
  \centerline{\includegraphics[scale=1]{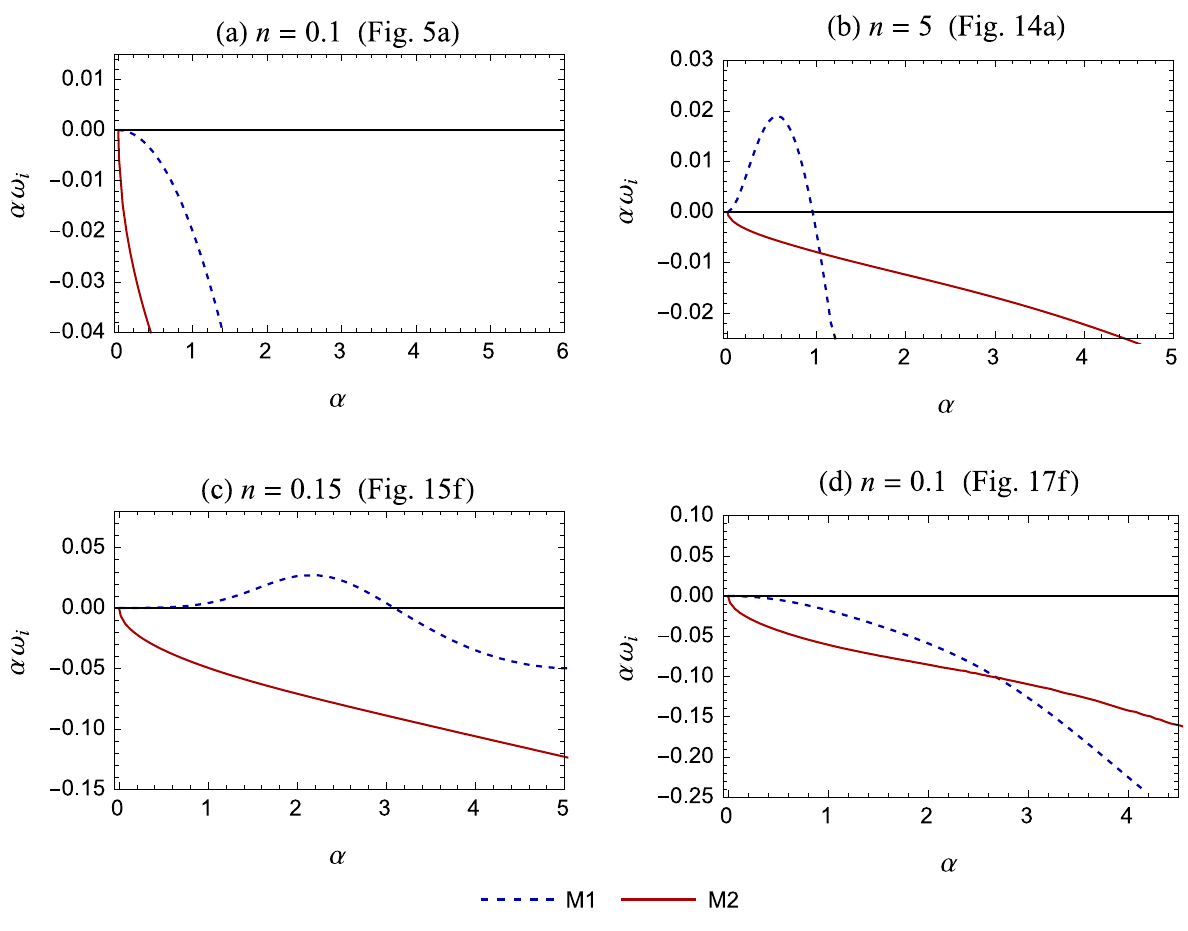}}
  \caption{Stabilization of short wave modes (M1-SW and M2-SW) when one fluid layer is much thinner than the other. Each plot has the same parameter values as a previous figure in this paper (mentioned above each plot), with the exception of the thickness ratio $n$. Moderate values of $n$ are used in the previous figures, wherein one or both of the short wave modes are unstable. In contrast, the new values of $n$, displayed above each plot, correspond to cases in which either fluid 1 or fluid 2 has a relatively small depth. Making one of the fluid layers thin (while $Ma$ is fixed) clearly has a stabilizing effect on the short wave modes (compare each plot with the figure mentioned above it).}
\label{fig:nextreme}
\end{figure}

Neither of the short wave instability modes (M1-SW, M2-SW) are identified by the asymptotic analysis of \citet{Wei2006}. Their absence in the thin layer limit implies that, for a fixed finite $Ma$, the short wave modes are stabilized as one fluid layer is made much thinner that the other. Numerical calculations that demonstrate this behavior are presented in Fig. \ref{fig:nextreme}. Each plot in this figure is a thin layer variation of a case studied in a previous section of this paper, wherein one or both of the short wave modes are unstable. For example, Fig. \ref{fig:nextreme}a is plotted for the same parameter values (including $Ma$) as Fig. \ref{fig:shortboth}a, except for the thickness ratio $n$. For $n=1$ (Fig. \ref{fig:shortboth}a) both M1-SW and M2-SW are unstable, but when the depth of fluid 2 is significantly decreased ($n = 0.1$, Fig. \ref{fig:nextreme}a) both short wave modes become stable. Comparing Fig. \ref{fig:nextreme}b with Fig. \ref{fig:flux_long}a shows that the M2-SW mode is stabilized on decreasing the thickness of fluid 1. The long wave mode M1-LW, on the other hand becomes more unstable as $n$ is increased, in accordance with the long wave asymptotic prediction \eqref{longexp}. Fig. \ref{fig:nextreme}c and Fig. \ref{fig:nextreme}d show that this behaviour persists when inertial effects are included (nonzero $\Rey$). Consequently, the simplified equations of the thin layer limit \citep{Wei2006} predict only long wave instability modes. An analysis of the governing equations for moderate thickness ratios will most likely uncover M1-SW and M2-SW modes in Couette flow as well.

Although \citet{Wei2006} did not observe the M1-SW and M2-SW modes, he did find a range of $Ca$ in which the long wave instability transitions to an intermediate wavelength mode that has a range of unstable wavenumbers bounded away from zero (cf. Fig. 4 of \citet{Wei2006}). This mode cannot be identified with the M1-SW mode because it was found to stabilize completely on decreasing $Ca$, which is contrary to the behaviour of the M1-SW mode (cf. \S \ref{sec:comp}). Moreover, since it transitions smoothly to the long wave instability on increasing $Ca$, it is not the M2-SW mode either. Instead it is most likely the analogue of the intermediate wavelength M1 mode shown in Fig. \ref{fig:transm1p5}d of \S \ref{sec:trans}, which arises as a distinct mode in the transition of M1-SW to M1-LW modes.

The physical problem studied by \citet{You2014a} is identical to that of the present work. However, they do not account for the effect of interface deformation on the concentration perturbations at the interface. As a result, the Marangoni stresses due to interface deformation are absent (the second term multiplying $Ma$ in \eqref{tanstress}). In this work, we have shown that these Marangoni stresses are the cause of the long wave Marangoni instability. In terms of the energy budget, the $MAS_I$ term that is dominant in the budget of the M1-LW mode is absent. Consequently, the M1-LW mode will not appear in the model analyzed by \citet{You2014a}. 

\citet{You2014a} do, however, report long wave instabilities at very small $\Rey$. These are unstable even when $n^2<m$, in contradiction with our asymptotic result \eqref{longexp} for long wave Marangoni modes. These modes do not correspond to the viscosity induced mode either, because \citet{You2014a} report that they are stabilized as $\Rey$ is increased beyond unity. Moreover, their calculations for cases of $n=1$ and $m \ne 1$ show the flow to be stable when $\Rey>1$. This is in contradiction with the established results of \cite{Yih1967} and \citet{yiantsios}, which predict instability to the viscosity induced mode at any non-zero value of $\Rey$ when $n=1$ and $m \ne 1$. On the other hand, both our asymptotic and numerical results are consistent with the instability of the viscosity induced mode at finite $\Rey$ \citep{yiantsios}.

\citet{You2014a} also report one short wave instability that occurs above a critical value of $Ma$, at small $\Rey$. Accounting for differences in the definition of dimensionless groups, we find that the neutral stability curves presented in \citet{You2014a} correspond to large $Ca$ of $O(10^3)$ and greater. Since the M2-SW mode dominates over the M1-SW mode at very large $Ca$ (cf. \S \ref{sec:comp}), the short wave mode reported by \citet{You2014a} is probably the M2-SW mode.

In summary, \citet{Wei2006} found the M1-LW mode in \textit{Couette} flow, while \citet{You2014a} found the M2-SW mode in \textit{Poiseuille} flow. In this work, we have found an additional short wave mode - M1-SW - that becomes unstable in the region where the long wave M1-LW mode is stable. The dominant instability is shown to switch between these three modes as parameters are varied. The M1-SW instability is dominant at smaller values of $Ca$ (cf. \S \ref{sec:comp}) and thus is expected to be important in microchannel flows.

\section{Conclusions}\label{sec:conclusion}

This work has shown that the presence of a soluble surfactant can destabilize stratified flow via solutal Marangoni effects, provided a transverse concentration gradient is maintained across the fluids. Three distinct Marangoni instability modes are present, which destabilize the system even in the limit of creeping flow. One of these is a long wave mode (M1-LW), which is destabilized by concentration variations due to deformation of the interface in the presence of a base transverse concentration gradient. The other two modes are short wave instabilities (M1-SW and M2-SW), which are amplified by the coupling between the disturbance flow and the interface concentration perturbations. One of the short wave modes (M1-SW) remains unstable even in the limit of large interfacial tension ($Ca \to 0$), wherein the interface is non-deforming. Thus, the base unidirectional flow may be unstable, leading to higher mass transfer rates, even in experiments which report a flat stationary interface.

When $\Rey$ is nonzero, the long wave viscosity-induced instability comes into play and interacts with the Marangoni instability. In certain regions of parameter space ($\gamma<1/K$, $n^2>m$, $m>1$), the viscosity-induced mode counteracts the long wave Marangoni mode and has a stabilizing influence. In other cases ($\gamma<1/K$, $n^2>m$, $m<1$), the viscosity-induced mode promotes the instability of long wave disturbances.

A summary of the instabilities that destabilize the system under different conditions is presented in Table \ref{tab:summary}. Surfactant laden creeping flow ($\Rey \ll 1$) is unstable to three different Marangoni instability modes. The long wave M1-LW mode is unstable when $n^2>m$ (table \ref{tab:summary}, (1)) whereas the short wave  M1-SW mode can be unstable when $n^2<m$ (table \ref{tab:summary}, (2)). The second short wave mode, M2-SW, can be unstable at any value of $n$ ((table \ref{tab:summary}, (1-2))). The short wave modes are unstable only if the magnitude of $Ma$ is greater than a critical value, which is different for each mode and depends on the other parameters. The critical $Ma$ is significantly greater when one of the fluid layers is much thinner than the other.

A necessary condition for these Marangoni instabilities is the presence of inter-fluid mass transfer in the base state. Therefore, the Marangoni modes are suppressed when equilibrium concentrations are maintained at the bounding plates ($\gamma = 1/K$) (table \ref{tab:summary}, (3)). The flow stability in this case is the same as that in the absence of surfactant effects ($Ma=0$). When $\Rey$ is increased to finite values the flow becomes unstable to the viscosity induced mode \citep{boomkamp2} (table \ref{tab:summary}, (4)). If soluble surfactant effects and inertia are present simultaneously, then both Marangoni and viscosity induced modes are present (table \ref{tab:summary}, (5)). The outcome of mutual interaction between these effects depends on the viscosity and thickness ratio of the fluids. Inertia can stabilize or destabilize the flow, as well as change the nature of the dominant instability from short wave to long wave, as shown in \S \ref{sec:inertia}.

The term ($n^2-m$) plays a key role in the stability characteristics of the M1-LW and M1-SW Marangoni modes (table \ref{tab:summary}). It is associated with the transverse gradient of the base state velocity field at the interface ($d\overline{u}_i/dy|_{y=0}$). The gradient is positive if ($n^2<m$), negative if ($n^2>m$) and zero if ($n^2=m$). This term is also prominent in the viscosity-induced mode (cf. \S \ref{sec:longwave} and \citet{yiantsios}) and in the instability caused by insoluble surfactants \citep{frenkel,halpern,Wei2005,Wei2007}. Elucidating the physical mechanisms through which this term influences these different instabilities is an interesting avenue for further work.

\begin{table}
  \begin{center}
\def~{\hphantom{0}}
  \begin{tabular}{lcccccc}
      Sr. No. & Condition  & M1-LW  & M1-SW & M2-SW & Viscosity-induced \\[3pt]
       1 & $\Rey \ll 1$, $n^2>m$   & \checkmark  &\ding{55} & \checkmark & \ding{55} \\
       2 & $\Rey \ll 1$, $n^2<m$   & \ding{55}  &\checkmark & \checkmark & \ding{55} \\
       3 & $\Rey \ll 1$, $\gamma=1/K$  & \ding{55}  &\ding{55} & \ding{55} & \ding{55} \\
       4 & $\Rey > 1$, $\gamma=1/K$  & \ding{55}  &\ding{55} & \ding{55} & \checkmark \\
       5 & $\Rey > 1$ ($\gamma \ne 1/K$)   & \checkmark  &\checkmark & \checkmark & \checkmark \\
  \end{tabular}
  \caption{Summary of the instabilities present in the system under different conditions. The three Marangoni modes are the long wave and short wave instabilities from the M1 eigenvalue branch (M1-LW and M1-SW respectively), the short wave instability from the M2 branch. The viscosity-induced mode is the long wave instability first identified by \citet{Yih1967}. $\Rey$ is restricted to small values in this classification. $Ma$ is non-zero and $\gamma < 1/K$ in all cases, unless explicitly stated otherwise. The latter inequality implies that mass transfer occurs from plate 1 to plate 2. Reversing the direction of mass transfer reverses the condition on the transition between long and short waves.}
  \label{tab:summary}
  \end{center}
\end{table}

An important task for future work is the extension of the present two-dimensional stability analysis to the analysis of three dimensional instability modes, i.e to include disturbances with variations in the direction perpendicular to the flow but parallel to the bounding plates. \citet{Wei2006} has shown that such 3D modes are more unstable than 2D modes for the case of thermocapillary instability in stratified Couette flow. It is thus quite possible that the inclusion of 3-D modes will lower the critical Marangoni number of the present Poiseuille stratified flow.

In an effort to understand the key effects of solutal-Marangoni stresses on the flow's stability, we have used a simplified model for mass transfer of the solute. In this model, adsorption and desorption of the solute to and from the interface is assumed to be instantaneous. This is an idealization. For a solute which adsorbs and desorbs at a finite rate, the distribution of solute at the interface will be affected by surface convection and diffusion in the Gibbs adsorption layer. These processes can have a subtle effect on the dynamics of the system. In the context of solutal Marangoni instability in \textit{stationary} fluid layers, accumulation and transport at the interface can stabilize or destabilize the system, depending on parameter values. The temporal nature of the linear instability modes (oscillatory or stationary) can also be affected \citep{Kovalchuk2006,Schwarzenberger2014}. In the nonlinear regime, surface transport of solute can lead to complex dynamical states, such as spontaneous oscillations \citep{Tadmouri2010}. It would therefore be interesting to extend the present model to account for these processes, and examine their influence on the stability of flowing fluid layers.

This study indicates that the solutal Marangoni instability can play an important role in applications involving low $\Rey$ stratified flow, such as solvent extraction in microchannels \citep{Assmann2013}. It has revealed the importance of the direction of mass-transfer between the fluids, which controls whether a long wave instability or a short wave instability is observed. This is of practical importance since long wave modes become unstable at any non-zero $Ma$ while short wave instability modes require $Ma$ to be greater than a critical value for instability. In order to accurately predict the instability threshold for these systems, however, the longitudinal variation of the base state concentration field must be accounted for. If this variation is gradual (large $\Pen$), then a weakly nonparallel stability analysis \citep{Huerre1990,Chomaz2005} can be carried out, in which the base state is treated as non-varying \textit{locally} at each point along the flow direction. In case the longitudinal concentration variation is rapid (small $\Pen$), then the disturbance cannot be decomposed into wave-like normal modes. Instead \textit{global} eigenmodes \citep{Chomaz2005,Theofilis2003}, with aperiodic variation along both the transverse and longitudinal directions, should be analyzed.\\

JRP would like to dedicate this paper to an inspiring teacher and close friend, Prof. Satyajit (Sat) Ghosh (VIT University, India and Leeds University, UK). The authors thank Ranga Narayanan for stimulating discussions and suggestions. JRP and TGR are grateful to MHRD, Gov. of India, for financial support.
\vfill

\appendix

\section{}\label{Appendix}
The expression for $G(n,m)$ used in the asymptotic solution \eqref{longexp} of \S \ref{sec:longwave} is given below: 
\begin{gather}
G(n,m)={m^7} + 6{m^7}n + 32{m^5}{n^2} + 344{m^5}{n^3} + (408{m^4} + 821{m^5}){n^4} \;\;\;\;\;\;\;\;\; \;\;\;\;\;\;\;\;\; \nonumber\\
 + (224{m^3} + 1642{m^4} + 1426{m^5}){n^5} + (1240{m^3} + 3667{m^4} + 1096{m^5}){n^6}  \;\;\;\;\;\;\;\;\;\;\;\;\;\;\;\nonumber\\
 + (224{m^2} + 3424{m^3} + 3424{m^4} + 224{m^5}){n^7} +(1096{m^2} + 3667{m^3} + 1240{m^4}){n^8} \nonumber\\ + (1426{m^2} + 1642{m^3} + 224{m^4}){n^9}
+(821{m^2} + 408{m^3}){n^{10}} + 344{m^2}{n^{11}}  \;\;\;\;\;\;\;\;\;\nonumber\\ + 32{m^2}{n^{12}} + 6{n^{13}} + {n^{14}}
-\bigg[2{m^6}n + 9{m^6}{n^2} + 120{m^6}{n^3} + 88{m^6}{n^4} \nonumber\\ \;\;\;\;\;\;\;\;\; \;\;\;\;\;\;\;\;\; \;\;\;\;\;\;\;\;\; \;\;\;\;\;\;\;\;\; \;\;\;\;\;\;
 + 88m{n^{10}} + 120m{n^{11}} + 9m{n^{12}} + 2m{n^{13}} \bigg] \label{Gexp}
\end{gather}

\bibliographystyle{jfm}
\bibliography{references2}

\end{document}